\documentclass[twocolumn]{aastex63}

\usepackage{natbib}

\usepackage{amsmath}
\usepackage{xcolor}
\shorttitle{Chondritic Solar Neighborhood}
\shortauthors{}

\graphicspath{{./Figures/}}

\begin{document}

\title{A Chondritic Solar Neighborhood}

\author{Isabella L. Trierweiler}

\affiliation{Department of Earth, Planetary, and Space Sciences, University of California, Los Angeles, Los Angeles, CA 90095, USA}

\correspondingauthor{Isabella L. Trierweiler}
\email{isabella.trierweiler@astro.ucla.edu}

\author{Alexandra E. Doyle}
\affiliation{Department of Earth, Planetary, and Space Sciences, University of California, Los Angeles, Los Angeles, CA 90095, USA}

\author{Edward D. Young}
\email{eyoung@epss.ucla.edu}
\affiliation{Department of Earth, Planetary, and Space Sciences, University of California, Los Angeles, Los Angeles, CA 90095, USA}


\begin{abstract}
A persistent question in exoplanet demographics is whether exoplanetary systems form from similar compositional building blocks to our own. Polluted white dwarf stars offer a unique way to address this question as they provide measurements of the bulk compositions of exoplanetary material. We present a statistical analysis of the rocks polluting oxygen-bearing white dwarfs and compare their compositions to rocks in the Solar System.  We find that the majority of the extrasolar rocks are consistent with the composition of typical chondrites. Measurement uncertainties prevent distinguishing between chondrites and bulk Earth, but do permit detecting the differences between chondritic compositions and basaltic or continental crust.  We find no evidence of crust amongst the polluted white dwarfs. We show that the chondritic nature of extrasolar rocks is also supported by the compositions of local stars. While galactic chemical evolution results in variations in the relative abundances of rock-forming elements spatially and temporally on galaxy-wide scales, the current sample of polluted white dwarfs are sufficiently young and close to  Earth that they are not  affected by this process.  We conclude that exotic compositions are not required to explain the majority of observed rock types around polluted white dwarfs, and that variations between exoplanetary compositions in the stellar neighborhood are generally not due to significant differences in the initial composition of protoplanetary disks.  Nonetheless, there is evidence from stellar observations that planets formed in the first several billion years in the Galaxy have lower metal core fractions compared with Earth on average. 
\end{abstract}

\keywords{White dwarf stars (1799)}


\section{Introduction}\label{section:intro}

The growing sample of exoplanets has inspired many studies detailing their compositions and interiors. Analyses of exoplanet compositions using mass and radius relationships or through extrapolating stellar abundances have led to  a wide range of possible exoplanet compositions \citep{Bond2010}, including Earth-like compositions, but also carbon-rich planets \citep[e.g.][]{Dorn2019}, coreless super-Earths \citep[e.g.][]{Madhusudhan2012}, and mineralogies with no Earth-rock counterparts \citep{Putirka2021}. This hypothesized diversity of exoplanet compositions motivates us to benchmark the variety of putative non-Earth like planets against the compositions of exoplanetary rocks accreted by polluted white dwarfs (WDs). The metal pollution on WDs is caused by accretion of exoplanetary debris and provides direct measurements of bulk compositions of extrasolar rocks that are not susceptible to the same degeneracies as the mass/radius approach \citep[e.g.][]{Dorn2015}. The vast majority of WD pollutants are rocky, with some fragments identified as specifically core or crust-like \citep[e.g.][]{Doyle2019, Harrison2018, Hollands2018, Melis2017, Jura2014}. Some water-rich objects have also been identified, with possible parent bodies including Kuiper Belt analogs or exomoons \citep[e.g.][]{Doyle2021, Klein2021, Hoskin2020, Xu2017, Raddi2015}.

We analyze the abundances from 31 oxygen-bearing polluted WDs. The presence of O, along with other major rock-forming elements such as Si, Mg, and Fe indicate that these WDs are accreting rocky material. We compare the abundances of the WD pollution to rocks throughout the Solar System, an approach motivated by previous WD studies \citep[e.g.][]{Doyle2023, Swan2019, Xu2013}. We also carry out the same analysis for local stars, as a proxy for protostellar disk environments and as a broad representation of the system's rocky planet compositions \citep[e.g.][]{Schulze2021}. For this purpose, we use the Hypatia catalog of stars, which includes elemental abundances for thousands of stars within $\sim500$ pc of the Sun \citep{Hinkel2014}. Throughout, we compare WD and stellar compositions to solar system rocks using a reduced chi-squared goodness-of-fit test.

While individual stars may show unusual amounts of particular elements, we find in this work that the majority of WD pollution is indistinguishable from chondrites in composition, when accounting for uncertainties in the measured abundances. The whole-rock compositions of CI chondrites are considered a proxy for the relative abundances of rock-forming elements of the Solar System, as they are the best compositional match to the Sun \citep[e.g.][]{Lodders2009, Anders1989}, and we use them here as representative of chondrites in general.

This paper is organized as follows. In Section \ref{section:methods} we outline the $\chi^2$ calculation used to test the goodness of fit of each set of abundances to CI chondrite. To demonstrate the method, we apply the $\chi^2$ test to Solar System rocks in Section \ref{section:SS}. We then carry out fits for the WD polluters in Section \ref{section:WDs} and for the Hypatia catalog stars in Section \ref{section:hypatia}. We discuss the impact of galactic chemical evolution on polluted WD and Hypatia compositions in Section \ref{section:discussion} and present our conclusions in Section \ref{section:conclusion}.

\section{Methods}\label{section:methods}
Throughout this work we compare observed abundances to the CI chondritic composition \citep{Lodders2019} by computing reduced $\chi^2$ values ($\chi^2_\nu$). Measurement uncertainties for the WDs are propagated using a Monte Carlo approach. Uncertainties for the Hypatia catalog stars are gathered from the catalog \citep{Hinkel2014}. For each star, we use the relative concentrations of Si, Fe, Al, Ca, Ni, and Cr, where available, all normalized to Mg. We do not include more volatile elements such as C, N, or O in the comparisons as we are primarily concerned with rock compositions in this work. Because a very diverse range of physical processes can vary volatile abundances during planet formation \citep[e.g.][]{Bonsor2021}, volatile abundances are not necessarily related to rock compositions. Excluding these elements therefore allows for more direct comparison of the underlying rock to Solar System samples.  Additionally, while O is a major element in rocks, its abundance is correlated with the other included rock-forming elements in oxides, providing further motivation to exclude it from the $\chi^2_\nu$ calculations. 

Starting with log abundances for each star and WD, we construct a random sample of abundances for each element assuming a normal distribution based on the reported logarithmic abundance ratios and their uncertainties. We then transform the distribution of logarithmic relative abundances to a distribution of number ratios for each element relative to Mg. The reported symmetric errors in the logs lead to asymmetric distributions in number ratios, so we select our assumed abundance ratios and uncertainties as the median, 16.5, and 83.5 percentiles from the distributions. Errors in ratios of elements are obtained by propagation of uncertainties in the individual elements using Monte Carlo sampling.  To address the asymmetric uncertainties in the ratios of elements arising from reported symmetric errors in logs of the ratios for both the WDs and Hypatia catalog stars, we use the following equation to calculate the $\chi^2$ goodness of fit for each element $i$ relative to Mg:

\begin{equation}
    \chi^2_i = \left(\frac{\delta_i}{\sigma_i}\right)^2 \left(1 - 2A \frac{\delta_i}{\sigma_i} + 5A^2 \left(\frac{\delta_i}{\sigma_i}\right)^2 \right),
\end{equation}
\noindent where $\delta_i$ is the difference between the observed and expected element ratio, $\sigma_i$ is the average of the upper and lower errors, and $A$ describes the asymmetry in the errors as $A = (\sigma_+ - \sigma_-)/(\sigma_+ + \sigma_-)$, where $\sigma_+$ and $\sigma_-$ are the asymmetrical measurement uncertainties for element $i$ \citep{Barlow2003}. To find the reduced $\chi^2$,  we sum over all elements and divide by the degrees of freedom, taken to be the number of elements (excluding Mg) measured for the given star.

We define the passing conditions (accepting the alternative hypothesis $H_{\rm a}$ that the rocks are chondritic) for the $\chi^2_\nu$ tests using the parameter $\alpha$, the probability of randomly obtaining a $\chi^2_\nu$ value greater than the one calculated for the observed abundances (e.g.\, the probability of incorrectly rejecting the null hypothesis, $H_0$, that the rocks are not chondritic). Following convention, we place the $\alpha$ limit at $0.05$, so that any stars identified as chondritic compositions must have a $\chi^2_\nu$ with an $\alpha < 0.05$, implying a $H_{\rm a}=1-\alpha$ probability that the correspondence with chondrite is not due to random chance. 

Because our sample sizes are very small, we must account for errors in the $\chi^2_\nu$ values. The error in $\chi^2_\nu$ can be approximated as $\sigma = \sqrt{2/n}$ \citep{Andrae2010}, where $n$ is the number of data points for a given star's composition. We therefore define the critical reduced chi-square values as $\chi^2_{\nu, \rm crit} = \chi^2_\nu (\alpha = 0.05) + 2 \sqrt{2/n}$, allowing for a $2\sigma$ error in $\chi^2_\nu$. These constraints give critical $\chi^2_\nu$ values of $\sim 3$ to 4, for $n$ from $3-6$ (excluding Mg), varying inversely with the number of elements observed for each star. For a given star, if the elements available define $\chi^2_\nu$ $\lesssim 3$ to 4, the data are taken as evidence for chondritic rocky parent bodies or planets.

In order to identify outliers in the elemental abundances for each WD and Hypatia star, we apply a Dixon's Q test \citep{Dean1951} with a confidence level of $95\%$ ($p=0.05$). We choose this test as it is best suited for small sample sizes. For this test we convert abundances to $(n_{\rm Z}/n_{\rm Mg}) / (n_{\rm Z}/n_{\rm Mg})_{\rm CI}$ such that 1 represents a perfect fit to chondrite. Outlier elements are therefore the elements with the worst fits to chondrite (other than Mg), and we identify an outlier in six of the WDs. Stars which pass as chondritic when an outlier is ignored are considered ``soft passes." 

\section{Solar System Rocks}\label{section:SS}
To test our ability to differentiate between different rock types using the methods of Section \ref{section:methods}, we first apply our test for chondritic compositions to rocks in the Solar System, including bulk Earth (BE) and bulk silicate Earth (BSE, \cite{McDonough2003}), mid-ocean ridge basalt (MORB), continental crust (CC, \cite{Rudnick2003}),  bulk silicate Mars (BSM, \cite{Taylor2013}), and E chondrites (EH, \cite{Wasson1988}).  For each element, we apply the mean uncertainty calculated from our sample of WDs for that element, with the resulting uncertainties generally ranging from 0.15$-$0.30 dex. 

\begin{figure}
\centering
    \includegraphics[width=0.5\textwidth]{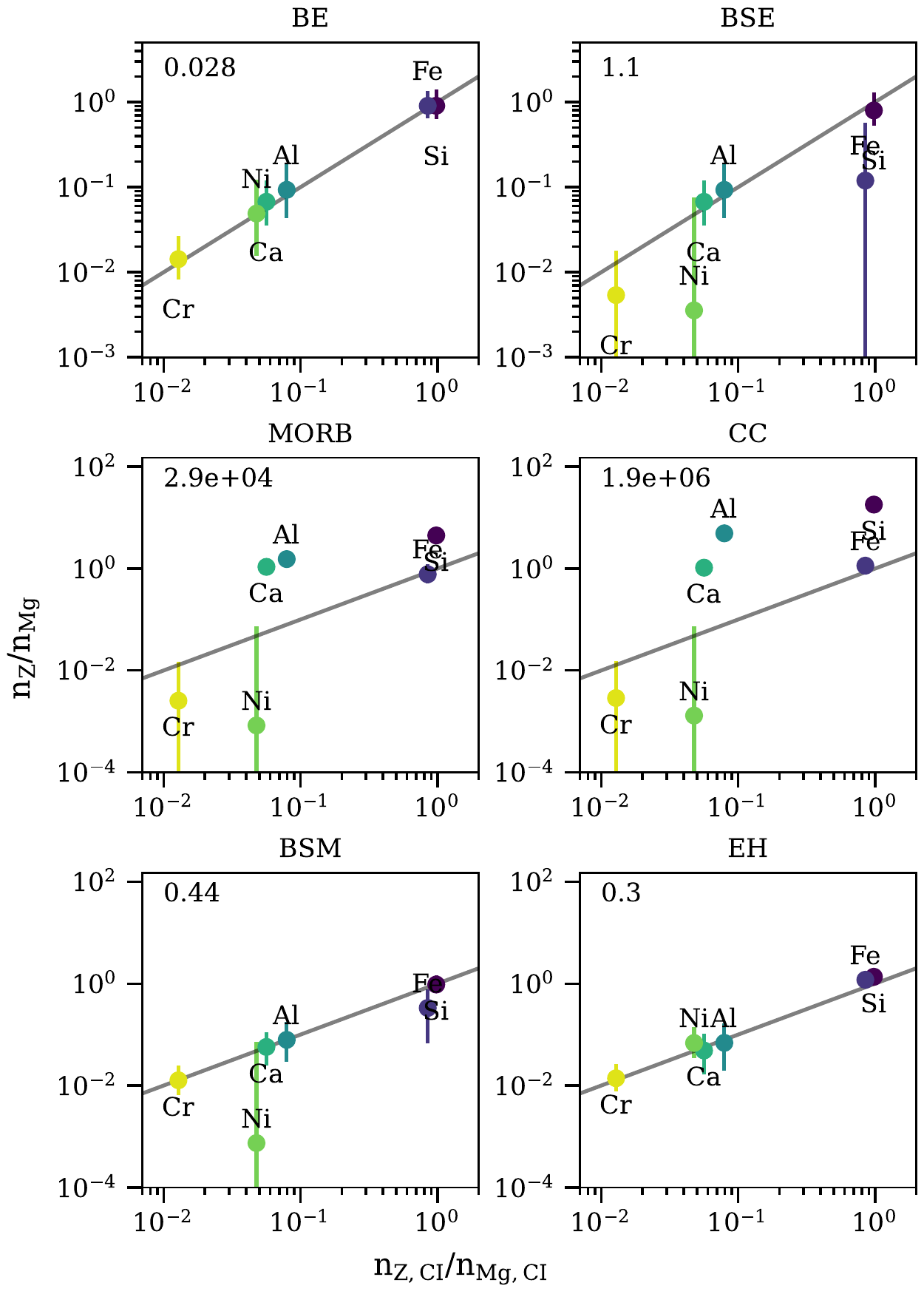}
\caption{Compositions of Solar System rocks compared to CI chondrite, for Bulk Earth (BE), Bulk Silicate Earth (BSE), Mid-Ocean Ridge Basalt (MORB), Continental Crust (CC), Bulk Silicate Mars (BSM), and E chondrites (EH). Error bars for each element correspond to the mean uncertainty in the WD abundances for that element. The reduced $\chi^2$ value for each correlation with CI chondrite is indicated in the plots at the upper left. Where $\chi^2_\nu$ $\lesssim 3$ to 4, the data are taken as evidence for chondritic rocky parent bodies or planets. }\label{Fig:Earthrocks_1-1}
\end{figure}

We find that BE, BSE, BSM, and the E chondrites are indistinguishable from CI chondrites, while MORB and CC are very clearly not good matches to CI chondrite in these tests (Figure \ref{Fig:Earthrocks_1-1}). Bulk Earth being indistinguishable from chondritic is in contrast with the distinction typically drawn between the two rock types in previous studies \citep[e.g.,][]{Putirka2019, Drake2002}, and is the result of propagating the large uncertainties associated with the element ratios for the WDs compared with the comparatively small differences in composition among the rock compositions. 

Throughout this work, we report compositions as consistent or inconsistent with chondrites, while recognizing that with current measurement uncertainties, chondrites, Earth, and Mars are all indistinguishable. However, this test is able to definitively differentiate between chondrite-like compositions and crust, the latter representing products of igneous differentiation of chondrites. 

\section{White Dwarfs}\label{section:WDs}
Our WD sample includes 31 WDs with detections of oxygen together with other rock-forming elements (Table \ref{Table:star_table}). The vast majority of the WDs in our sample have atmospheres that are helium-dominated, and a handful are hydrogen-dominated. The WDs are all within about $200 \rm pc$ of the Sun. For each WD, we draw stellar properties, elemental abundances  and uncertainties in abundances from the references listed in Table \ref{Table:star_table}, supplemented by the Montreal White Dwarf Database (MWDD, \citealt{Dufour2017}). We use the elements Si, Fe, Al, Ca, Ni, and Cr where available. We ratio all abundances to Mg and propagate uncertainties using a Monte Carlo approach outlined in Section \ref{section:methods}.

\begin{table*}
\caption{All WD parameters are collected from the references listed in the table. Any values not reported by the paper have been supplemented using the Montreal White Dwarf Database \citep{Dufour2017}. Throughout this work we group WDs by the dominant element in their atmospheres (H or He-dominated only). }
\begin{center}
\begin{tabular}{l|c|c|c|c|c|c}
\hline
White Dwarf* & Type & $\rm T_{\rm eff}$ (K)  & log(g) & D (pc) & M$^\dagger (\rm M_\odot)$  & Reference \\ \hline 
G29-38 & H & 11820 & 8.15 & 14 & 0.696 & \cite{Xu2014}\\
SDSSJ1043+0855 (SDSS J104341.53+085558.2) & H & 18330 & 8.05 & 169 & 0.649 & \cite{Melis2017}\\
PG1015+161 & H & 19226 & 8.04 & 88	& 0.645 & \cite{Gansicke2012}\\
WD1226+110 & H & 20900 & 8.15 & 129	&	0.714 & \cite{Gansicke2012}\\
WD1929+012 (GALEX J193156.8+011745) & H & 21200 & 7.91 & 53	&	0.578 & \cite{Gansicke2012}\\
PG0843+516 (WD 0843+516) & H & 23095 & 8.17 & 139	&	0.730 & \cite{Gansicke2012}\\
WD0446-255 & He & 10120 & 8.00 & 91	&	0.581 & \cite{Swan2019}\\
WD1350-162 & He & 11640 & 8.02 & 108 &	0.596 & \cite{Swan2019}\\
WD1232+563 & He & 11787 & 8.30 & 172	&	0.773 & \cite{Xu2019}\\
SDSSJ1242+5226 (SBSS 1240+527)& He & 13000 & 8.00 & 161	&	0.587 & \cite{Raddi2015}\\
SDSSJ2339-0424 (GALEX J233917.0-042425) & He & 13735 & 7.93 & 89 &	0.548 & \cite{Klein2021}\\
SDSSJ0738+1835 (SDSS J073842.56+183509.6) & He & 13950 & 8.40 & 172	&	0.842 & \cite{Dufour2012}\\
HS2253+8023 & He & 14000 & 8.10 & 72	&	0.648 & \cite{Klein2011}\\
WD1425+540 & He & 14490 & 7.95 & 52 &	0.560 & \cite{Xu2017}\\
WD1145+017 & He & 14500 & 8.11 & 146 &	0.655 & \cite{Fortin-Archambault2020}\\
GaiaJ0218+3625 (GALEX J021816.6+362507) & He & 14691 & 7.86 & 116	&	0.512 & \cite{Doyle2023}\\
EC22211-2525 & He & 14743 & 7.90 & 109 &	0.534 & \cite{Doyle2023}\\
WD2207+121 & He & 14752 & 7.97 & 164 &	0.572 & \cite{Xu2019}\\
WD1551+175 & He & 14756 & 8.02 & 162 &	0.601 & \cite{Xu2019}\\
WD1244+498 & He & 15150 & 7.97 & 164	&	0.573 & \cite{Doyle2023}\\
WD1248+1004 (SDSS J124810.23+100541.1) & He & 15178 & 8.11 & 73	&	0.656 & \cite{Doyle2023}\\
GD40 & He & 15300 & 8.00 & 120	&	0.591 & \cite{Jura2012}\\
G241-6 & He & 15300 & 8.00 & 65 &	0.591 & \cite{Jura2012}\\
GaiaJ1922+4709 (Gaia DR2 2127665711125011456) & He & 15497 & 7.95 & 127 &	0.562 & \cite{Doyle2023}\\
GD378 & He & 15620 & 7.93 & 44	&	0.551 & \cite{Klein2021}\\
SDSSJ1734+6052 (GALEX J173435.7+605203) & He & 16340 & 8.04 & 150	&	0.616 & \cite{Doyle2023}\\
GD61 & He & 17280 & 8.20 & 54 &	0.715 & \cite{Farihi2013}\\
WD1415+234 & He & 17312 & 8.17 & 127 &	0.696 & \cite{Doyle2023}\\
SDSSJ2248+2632 (SDSS J224840.97+263251.7) & He & 17369 & 8.02 & 123 &	0.606 & \cite{Doyle2023}\\
Ton345 & He & 19780 & 8.18 & 106	&	0.706 & \cite{Wilson2015}\\
WD1536+520 & He & 20800 & 7.96 & 201	&	0.578 & \cite{Farihi2016}
\end{tabular} 
\end{center} *Where applicable, alternate WD identifiers are listed in parentheses. \\ $\dagger$ Masses are collected from the MWDD using the $\rm T_{eff}$ and $\rm log(g)$ values in the table. \label{Table:star_table} 
\end{table*}

We analyze both the raw and steady-state adjusted abundances for the WD pollution. The steady-state adjustment accounts for differential settling rates for different elements in the atmosphere of a WD. Settling rates also depend on the dominate element in the atmosphere of the WD, and range from days to millions of years \citep{Koester2009, Blouin2018}. The steady-state settling factor we use is $\rm ({n_Z}/n_{Mg})_{SS} = ({n_Z}/n_{Mg}) \times (\tau_{Mg}/\tau_{Z})$, where $\tau_{\rm Mg}$ and $\tau_{\rm Z}$ are the settling timescales for Mg and a given element Z, respectively. Settling timescales for the WDs in our sample are collected from the MWDD, using the WD parameters listed in Table \ref{Table:star_table}. These adjustments are clearly necessary  for the H-dominated WDs, where  settling is generally much more rapid. Suitability of the adjustment to abundances in the He-dominated atmospheres is less clear. We note that the stated steady-state factor is a simplistic approach to account for settling, which does not account for potential effects such as mixing in the WD atmosphere \citep[e.g.][]{Bauer2019, Cunningham2019}

\begin{figure*}
\centering
    \includegraphics[width=0.90\textwidth]{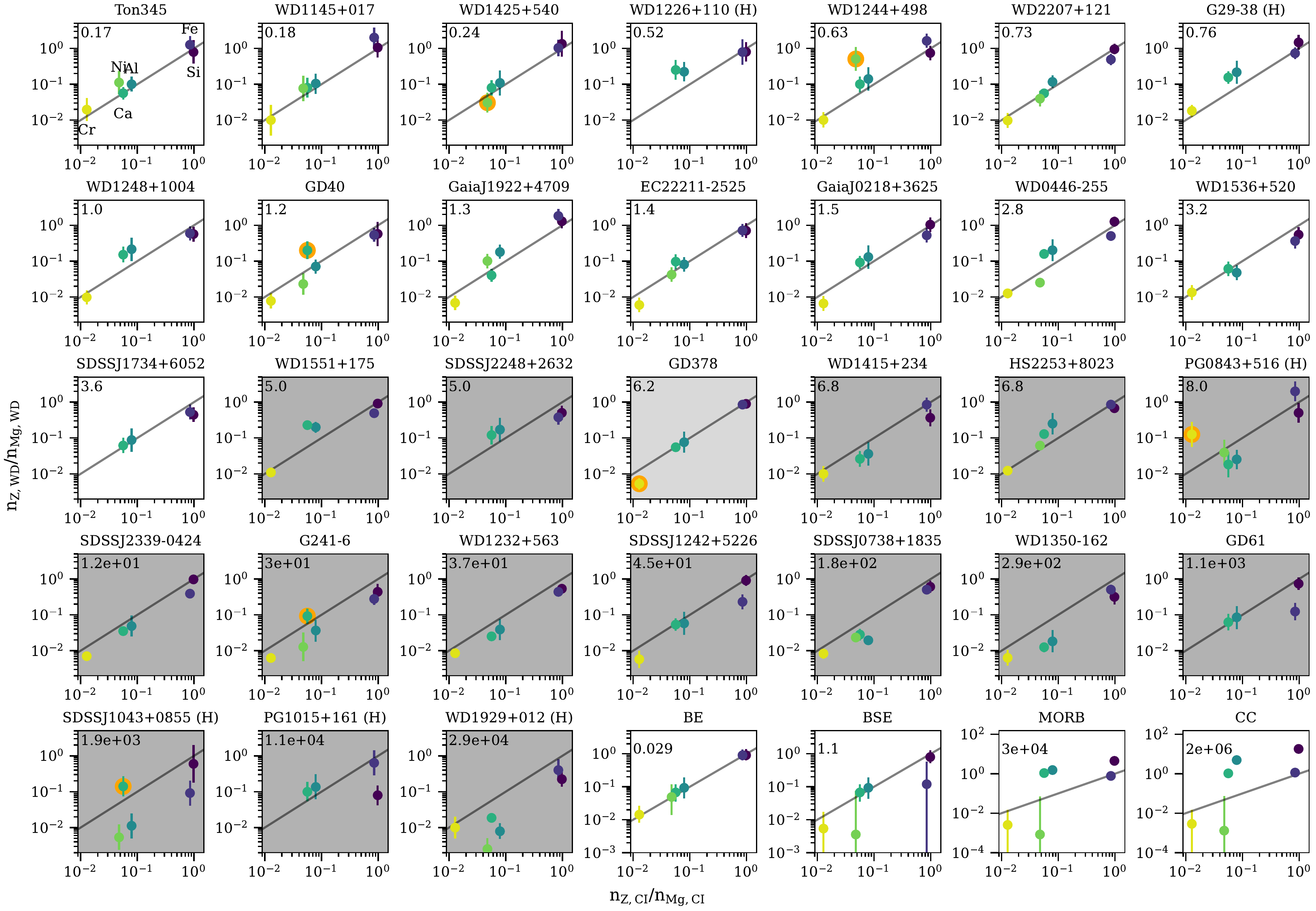}    
    \includegraphics[width=0.90\textwidth]{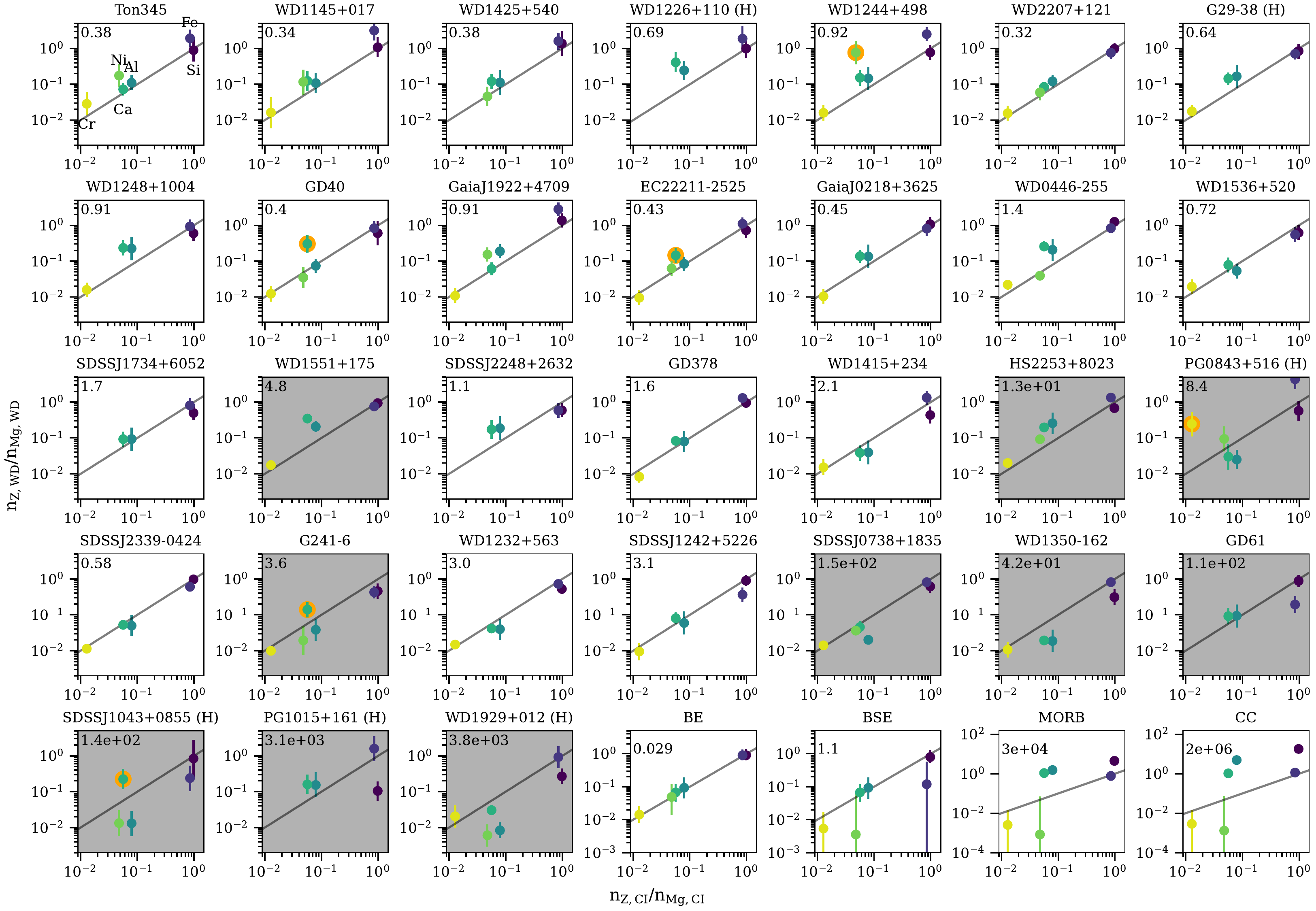}
   \caption{Comparison of WD pollution compositions to the CI chondrite composition, for the raw data (top) and steady-state adjusted abundances (bottom). The order of elements in each plot, from left to right, is Cr, Ni, Ca, Al, Fe, Si, all ratioed to Mg. The $\chi^2_\nu$ parameter appears in the upper left of each panel. WDs with white backgrounds are consistent with having accreted a chondritic rock composition.  WDs with dark grey backgrounds are not considered a good fit to chondrite, using an $\alpha$ parameter of 0.05 for goodness of fit. WDs with light grey backgrounds have an outlier element that allows the WDs to pass as chondritic when the outlier is removed (outliers highlighted in orange). }\label{Fig:O_WD_1-1}
\end{figure*}

For each WD, we compare the abundances normalized to Mg to the abundances measured in CI chondrites following the method outlined in Section \ref{section:methods}. Figure \ref{Fig:O_WD_1-1} shows this comparison for all WDs in our sample and for both the raw (top) and steady-state adjusted values (bottom). Hydrogen-dominated WDs are marked with ``H". From left to right on each plot, the element ratios are Cr/Mg, Ni/Mg, Ca/Mg, Al/Mg, Fe/Mg, and Si/Mg. The dark grey panels in Figure \ref{Fig:O_WD_1-1} indicate WDs which do not pass the $\chi^2_\nu$ test for chondritic composition. The lighter shaded panels show the ``soft pass" WDs, where ignoring an identified outlier allows the WD to pass as chondritic (see Section \ref{section:methods}). Solar system rocks are shown for comparison (see Section \ref{section:SS} for discussion of Solar System fits). 

\begin{figure}
\centering
    \includegraphics[width=0.5\textwidth]{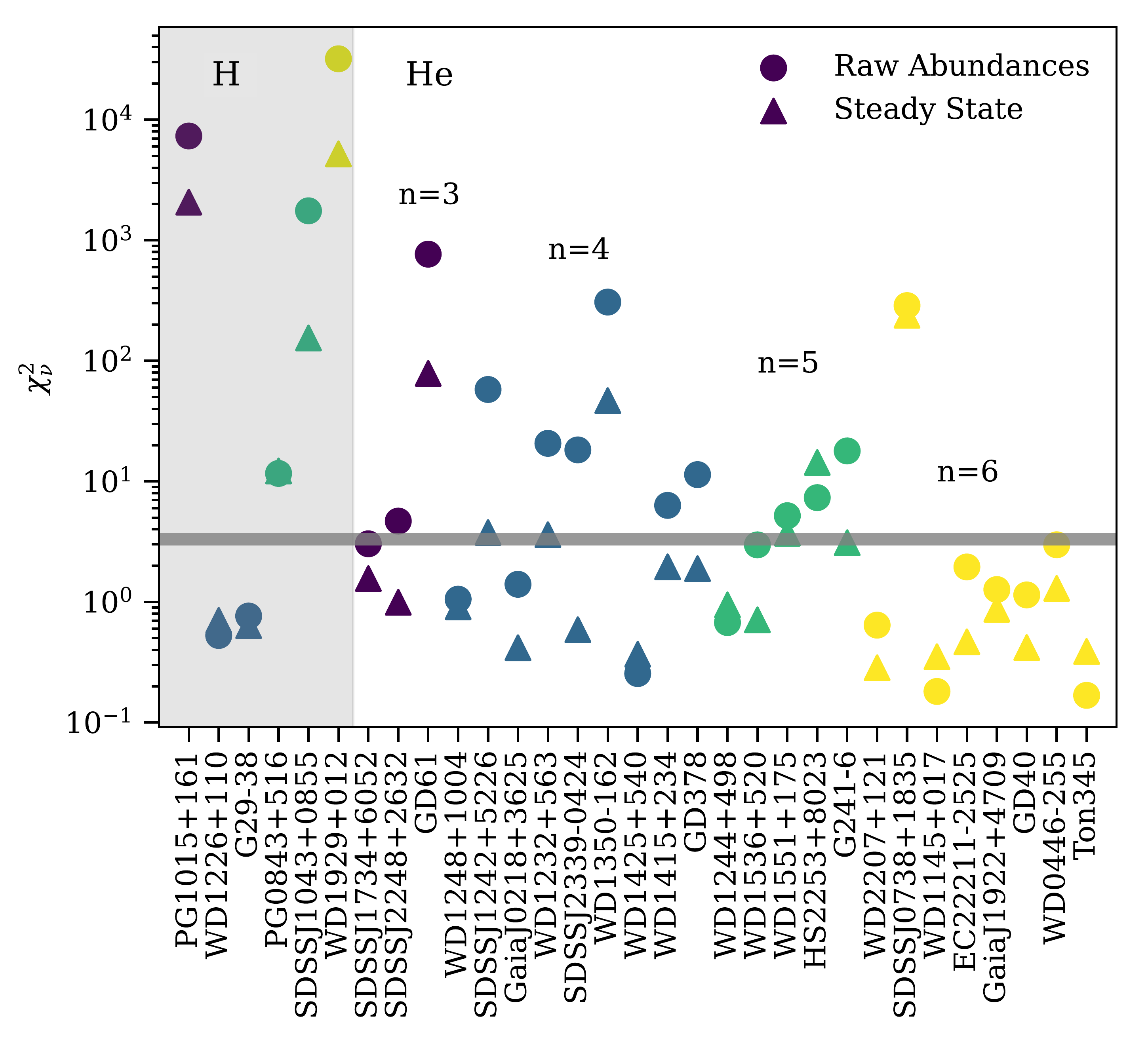}
   \caption{$\chi^2_\nu$ for each WD, relative to CI chondrite, for the raw abundances (circles) and steady-state adjusted values (triangles). The points are colored and grouped by $n$, the number of elements used to calculate $\chi^2_\nu$. The horizontal line shows a typical critical $\chi^2_\nu$ value ($\chi^2_\nu \sim 3$ to 4, based on the number of observed elements). In most cases, the steady-state values provide a better fit.}\label{Fig:WD_chi2}
\end{figure}

Figure \ref{Fig:WD_chi2} shows the $\chi^2_\nu$ parameters for the WDs for the raw data versus steady-state abundances, separated by the dominant element in the WD atmosphere. We also group the WDs by the number of observed elements considered in the statistical comparison ($n$), to illustrate the dependence of $\chi^2_\nu$ on $n$. Increasing $n$ generally lowers both the calculated and critical $\chi^2_\nu$ values. The condition for passing as chondritic at $n=3$ is $\chi^2_\nu \sim 4.2$ and at $n=6$ is $\chi^2_\nu \sim 3.3$.

We find that 15 of the 31 WDs pass the $\chi^2_\nu$ test as good matches to chondritic composition when using the raw abundances.  One additional WD passes as chondritic when its outlier element is ignored. A larger fraction of pollution passes as chondritic with the steady-state adjustment (21/31 pass). Because the steady-state adjustment does not improve the fits for every WD (Figure \ref{Fig:WD_chi2}), some WDs that pass as chondritic using the raw data do not pass in the steady-state case. We note that a larger proportion of WDs passing as chondritic in the steady-state case does not {\it a priori} mean the WDs are most likely to  be in the steady state phase of accretion. In any case, over half of the WDs in the sample are consistent with chondritic compositions using either the raw or steady-state compositions.

We find no compelling evidence for basaltic crust (MORB)  or continental crust rocks among the polluted WDs. When carrying out the same $\chi^2_\nu$ calculation for each WD relative to the other Solar System rock types considered here (Section \ref{section:SS}), no WDs are better fit by MORB or continental crust relative to CI chondrite, even those with $\chi^2_\nu$ values relative to chondrite of 100 and greater.

\subsection{White Dwarf Mineralogy Classification}
In addition to the $\chi^2_\nu$ test, we also follow the common practice of representing rock chemistries as ``normative mineralogies" in which elemental concentrations are converted to volumetric fractions of fictive minerals (\citealt{Cross1902}, see also \citealt{Putirka2021} Supplement for details).  We recast the WD pollution by projecting the observed abundances to a normative mineralogy composed of the relative abundances of Mg-endmember Olivine (OLV), Orthopyroxene (OPX) and Clinopyroxene (CPX). These minerals comprise a reasonable normative mineralogy used to classify ultramafic (e.g., peridotite) rocks, and chondrites are broadly similar to ultramafic rocks.  The fractions of these minerals in terms of moles depend on the relative numbers of Mg, Si and Ca atoms comprising the rocks. By inverting the mineral formulae for these reference minerals where OLV = Mg$_2$SiO$_4$, OPX = Mg$_2$Si$_2$O$_6$, and CPX = CaMgSi$_2$O$_6$, one obtains the function that transforms relative atomic abundances of Mg, Si, and Ca to the relative molar abundances of the minerals, which in matrix form is

\begin{equation}
    \begin{bmatrix}
    n_{\rm OLV} \\
    n_{\rm OPX} \\
    n_{\rm CPX} 
    \end{bmatrix} = 
    \begin{bmatrix}
    1 & -1 & 1 \\
    -1 & 2 & 3 \\
    0 & 0 & 1 
    \end{bmatrix} \times 
    \begin{bmatrix}
    n_{\rm Mg} \\
    n_{\rm Si} \\
    n_{\rm Ca}
    \end{bmatrix}.
\end{equation}
The molar abundances of the normative minerals are converted to approximate volume fractions (as is common for reporting rock mineralogies) using nominal molar volumes for OLV, OPX, and CPX, or $4.37$, $6.26$, and $6.60$ J/bar (J/bar  $= 0.1$ cm$^3$/mole). Fe and other less abundant elements are not included in this projection. Including Fe in this projection shifts the positions of the data somewhat, but does not substantially change the results.  


\begin{figure}
\centering
    \includegraphics[width=0.49\textwidth]{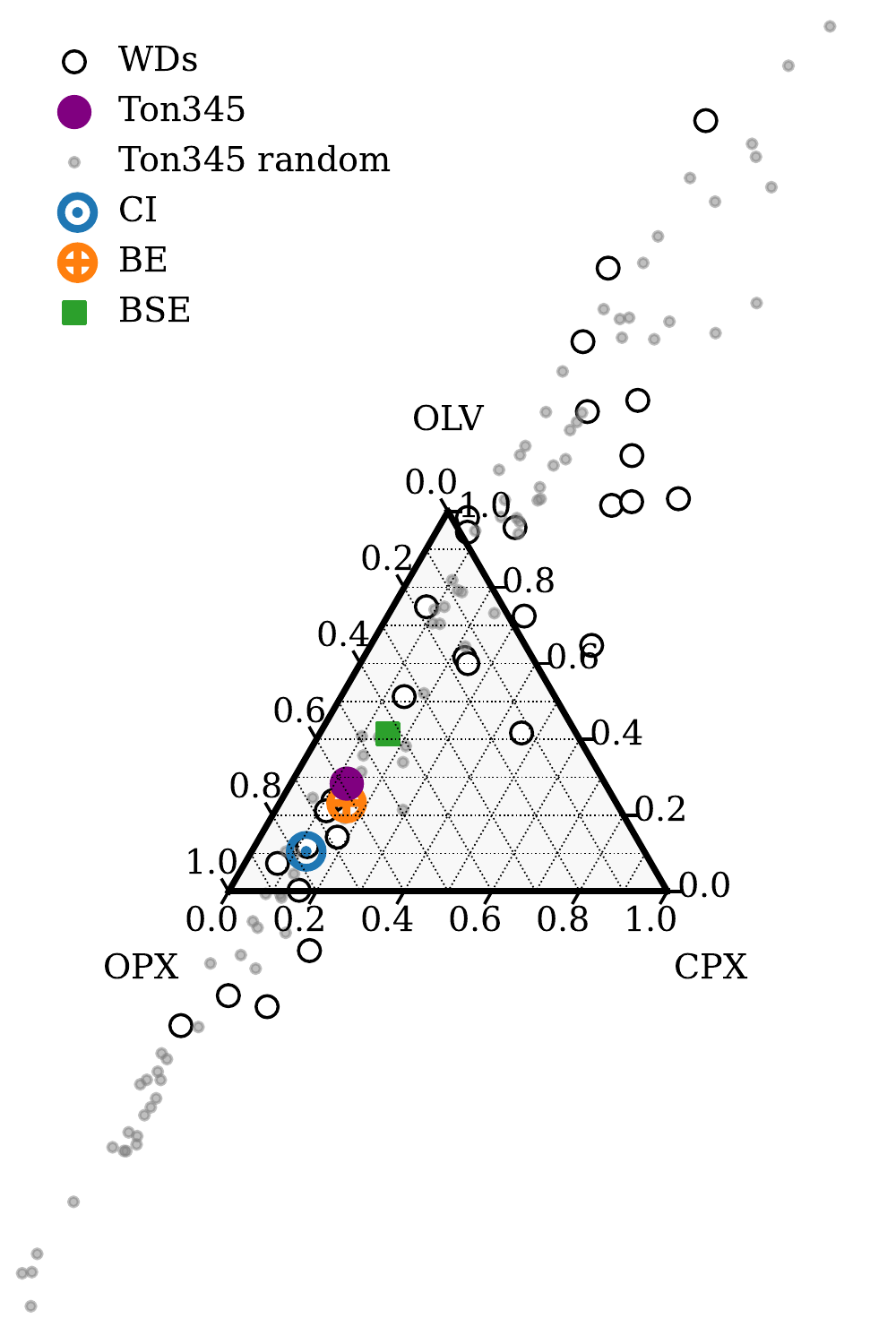}
   \caption{Ternary diagram for all of the WD samples, for raw abundances. The white points indicate the OLV, OPX, and CPX quantities derived from the median values of Mg, Ca, and Si for each WD. We also demonstrate the spread in OLV, OPX, and CPX that is due to the uncertainties in the WD abundances by showing the spread from 100 random draws of Mg, Si, and Ca for WD Ton345. The total spread in points is truncated for visibility. }\label{Fig:O_WD_tern}
\end{figure}

Figure \ref{Fig:O_WD_tern} shows the WD pollution represented as the relative volume fractions of OLV, OPX, and CPX implied by each composition. For each polluted WD, we take Monte Carlo draws of Mg, Si, and Ca using the reported values and corresponding uncertainties as the parent populations, and calculate the resulting normative mineral abundances. CPX is constrained only by the relative amount of Ca in the pollution, and exhibits comparatively little scatter as Ca uncertainties are generally small. We note that the volumetric fractions resulting from this method are not necessarily physical. Because this is a projection, some of the WD abundances result in negative amounts of OLV, OPX, or CPX, leading to scatter beyond the bounds of the positive ternary coordinate system. \cite{Putirka2021}  previously used this method to report exotic mineralogies for WD pollution, however we find that the uncertainties in Si and Mg are sufficiently large as to produce hopelessly large spreads in OLV and OPX abundances, so that it is impossible to constrain the mineralogy of the implied rocks (Figure \ref{Fig:O_WD_tern}). A similar spread in mineral abundances is derived from the steady-state data. We therefore conclude that categorizing rock pollution in WDs into rock types based on normative abundances of OLV, OPX, and CPX abundances, or similar normative mineralogies, is not possible. 

\section{Hypatia Catalog Stars}\label{section:hypatia}
Our Solar System exhibits a diversity of rock types originating from the same protoplanetary material, underscoring that samplings of rock can end up with very different compositions relative to the average starting material (e.g., crust vs.\ chondrites in Section \ref{section:SS}). To benchmark the ``final" exoplanetary rocks sampled by polluted WDs against protoplanetary material, we analyze the abundances of rock-forming elements in nearby stars by applying our compositional fitting method to stars in the Hypatia catalog \citep{Hinkel2014}. These stars should reflect protoplanetary material, to the extent that stellar abundances have been shown to broadly reflect compositions of planets around their stars  \citep[e.g.][]{Thiabaud2015, Bonsor2021, Schulze2021}. The stellar sample therefore represents a potential average of planet building materials, rather than the final rock compositions of individual rocky parent bodies sampled by the WDs.


We select Hypatia catalog stars with Mg and at least two other elements among Si, Fe, Al, Ca, Ni, or Cr. All uncertainties are obtained directly from the catalog, where they are listed as either the uncertainty reported in the original study or the mean uncertainty of multiple studies, where stars are observed by multiple methods. 

\begin{figure}
\centering
    \includegraphics[width=0.5\textwidth]{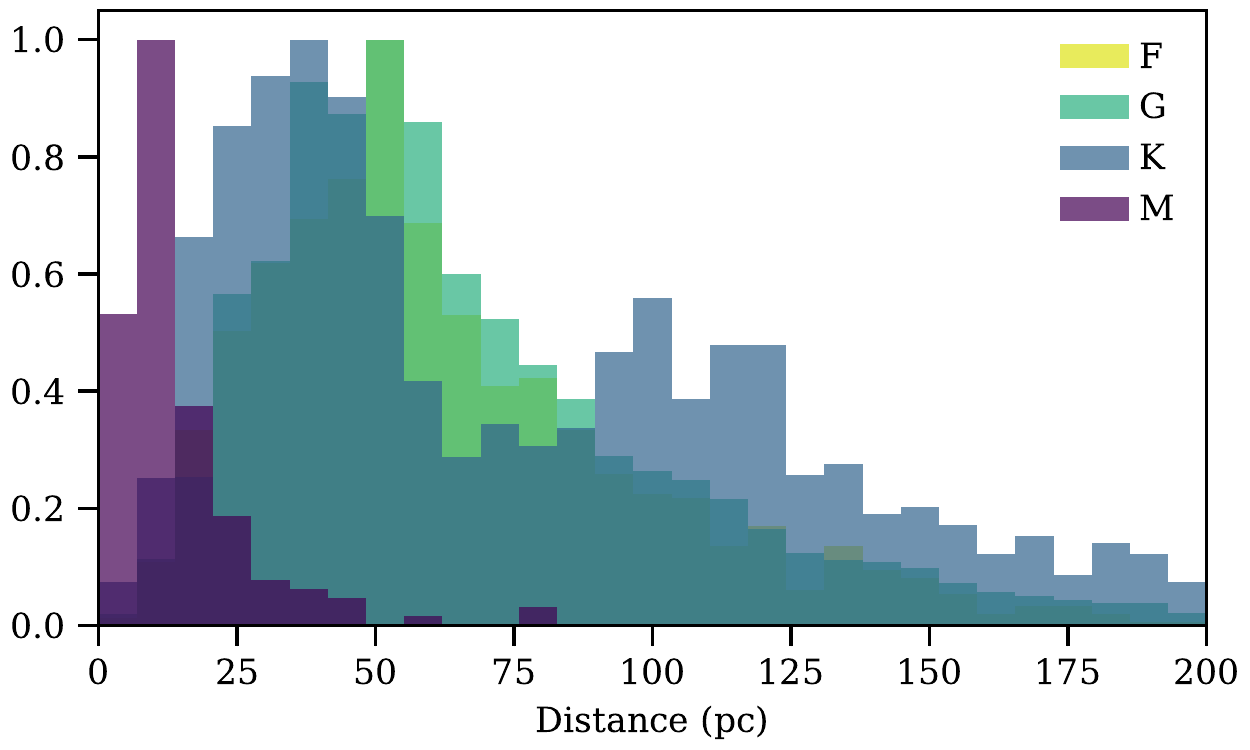} 
   \caption{Distribution of the distances of Hypatia catalog stars from the Sun (classified by stellar type). M dwarfs in the sample tend to be much closer to the Sun than other stars in the catalog. K stars also appear to have a bimodal distribution in distance.}\label{Fig:H_Mdist_bytype}
\end{figure}

Given the range of stars included in the Hypatia catalog, we explore how stellar type and distance may impact overall abundances. About 6500 stars in the catalog are classified as F, G, K, or M stars. In Figure \ref{Fig:H_Mdist_bytype}, we show the range of distances from the Sun in each classification. M stars in the sample tend to be much closer to the Sun $(<\sim 50 \rm pc)$ compared to the rest of the Hypatia stars. 

\begin{figure}
\centering
    \includegraphics[width=0.5\textwidth]{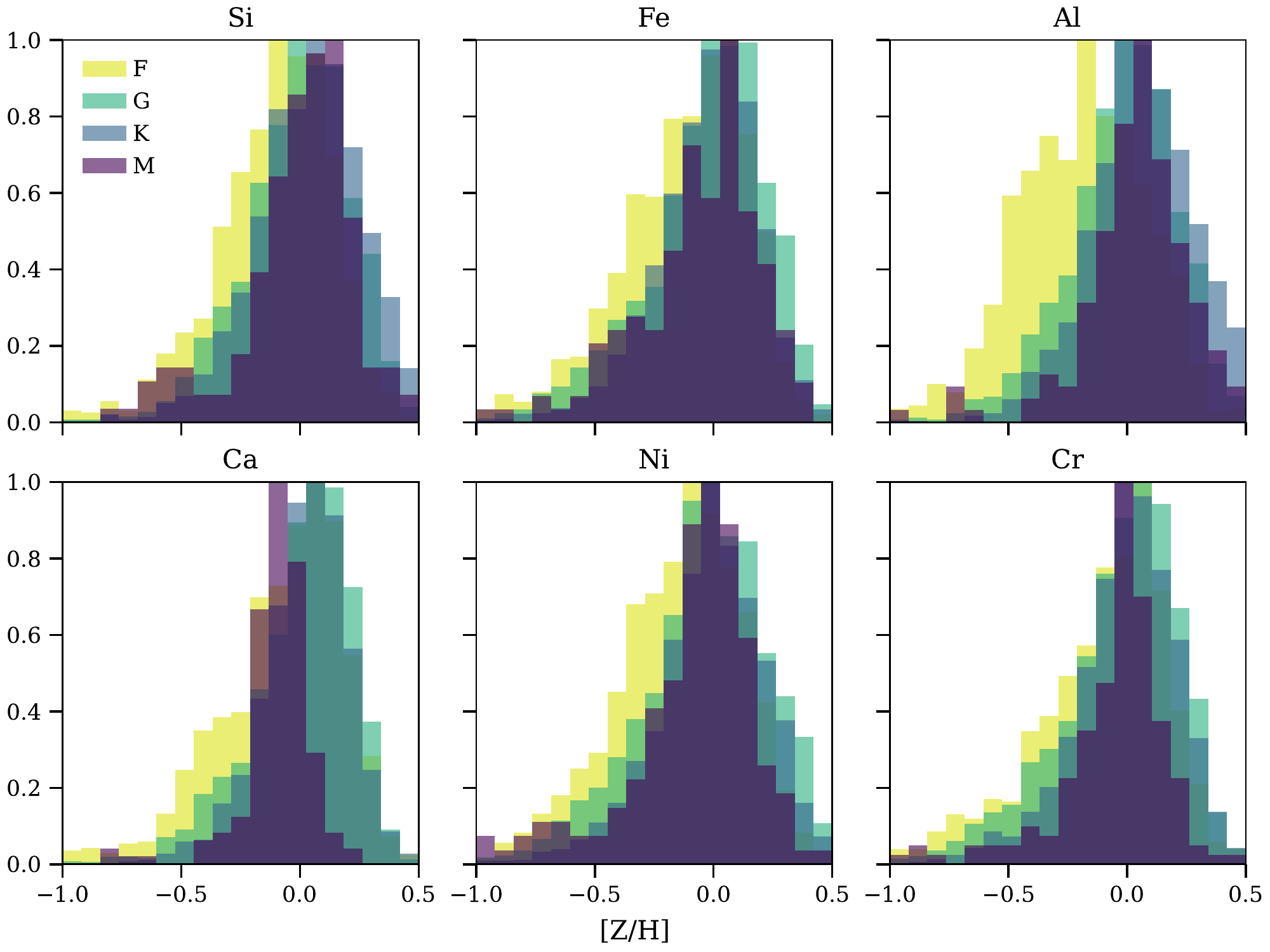} 
   \caption{Distribution of elemental abundances in Hypatia catalog stars relative to solar abundances, grouped by stellar type. All element distributions are well centered at solar abundances, with the exception of Ca in M stars and Al in F stars. }\label{Fig:H_Zdist_bytype}
\end{figure}

For the purposes of this work, we do not attempt to fully account for potential biases in the Hypatia catalog stars arising from the number of separate stellar surveys included in the catalog, but instead point out a few factors that are relevant to our compositional tests.  First, in Figure \ref{Fig:H_Zdist_bytype} we plot the distributions of elemental abundances relative to solar abundances, colored by stellar type. In general we find the distributions are centered around solar abundances, however we note a peak in Ca in M stars at lower abundances relative to other stellar types as well as a larger fraction of F stars with low Al than other stellar types. \cite{Hinkel2014} point out potential biases for both of these elements, including a lack of Al abundance measurements at higher metallicities, which may be altering the distribution. Additionally, most of the low [Ca/H] stars were drawn from the same single survey which may be inducing a spurious, non-physical bias in the [Ca/H] abundances. 

We also note that abundance uncertainties in the Hypatia catalog are strongly peaked at about 0.05 dex. Distance appears to have a strong influence on the uncertainties, with a larger range of uncertainties for stars closer to the Sun, though it is unclear if this is a physical effect or due to the stellar samples included. Stars within about 500 pc have a large range of uncertainties, up to 1.75 dex, while stars that are farther away have a nearly flat distribution of uncertainties at around 0.05 dex.  

From the Hypatia catalog we obtain abundances relative to solar abundances for each element, in the form $\rm [Z/H]= \log_{10}(Z/H)_{*} - \log_{10}(Z/H)_{\odot}$. We convert these relative abundances to molar ratios using the following equation:

\begin{equation}
    \rm \frac{n_Z}{n_{Mg}} = \frac{10^{[Z/H] + log_{10}(Z/H)_{\odot}}}{10^{[Mg/H] + log_{10}(Mg/H)_{\odot}}} = \frac{10^{[Z/H] + A(Z)}}{10^{[Mg/H] + A(Mg)}},
\end{equation}

\noindent where $\rm A(Z) = \log_{10}(Z/H)_{\odot} + 12$ is the solar abundance of the element $Z$, as defined in \cite{Lodders2009}. Uncertainties in the stellar abundances are propagated through this conversion using a Monte Carlo approach.

We calculate the $\chi^2_\nu$ goodness of fit parameter for Si, Fe, Al, Ca, Ni, and Cr, where available in each of the Hypatia catalog stars. For the elements considered in this work, we find median uncertainties of $\sim 0.05$ dex for the raw abundances relative to solar. To avoid invalid values for $\chi^2_\nu$, we replace any uncertainties of 0 with the median uncertainty for the corresponding element. 

\begin{figure*}
\centering
    \includegraphics[width=0.9\textwidth]{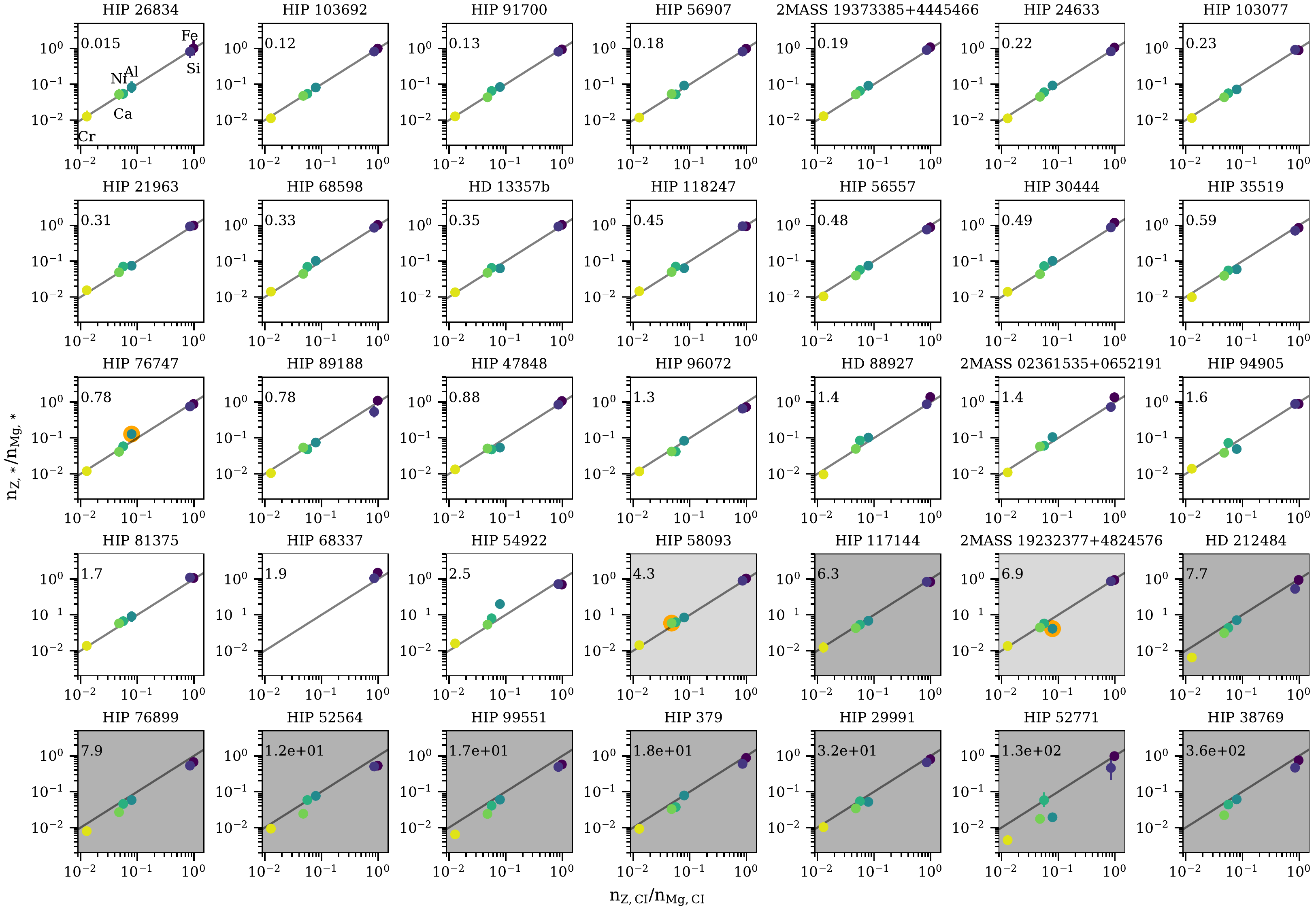}    
   \caption{Comparison of Hypatia catalog metal abundances to chondritic composition, for the 35 randomly chosen Hypatia catalog stars. The $\chi^2_\nu$ parameter is listed for each star. The order of elements in each plot, from left to right, is Cr, Ni, Ca, Al, Fe, Si, all ratioed to Mg. Stars with dark grey backgrounds are not considered a good fit to chondrite, using an $\alpha$ parameter of 0.05 for goodness of fit. Stars with light grey backgrounds are not a good fit for chondrite when all elements are used to calculate $\chi^2_\nu$, but do pass as chondritic when an outlier element is ignored (outliers highlighted in orange). }\label{Fig:H_1-1}
\end{figure*}

Figure \ref{Fig:H_1-1} shows the abundances for 35 randomly selected Hypatia stars. As with the WDs, white panels indicate stars that pass as chondritic, light grey panels show stars that pass when an outlier is ignored, and dark grey panels do not pass as chondritic even if outliers are ignored. We find that outliers do not make a big difference, and that about $75\%$ of stars pass as chondritic whether or not outliers are ignored. Similar to the WDs, we find that many of the stars that do not pass as chondritic are high in Mg, so that the abundances fall systematically below chondritic values. 

Because the uncertainties of the abundances vary strongly with the distance of each Hypatia star, we also compute fractions of chondritic stars considering only stars within 150 pc. We find that the results for the truncated sample are very similar to those for the full sample, with about $74\%$ of stars providing good matches to chondrites.

\subsection{Hypatia Mineralogy Classification}
\begin{figure}
\centering
    \includegraphics[width=0.49\textwidth]{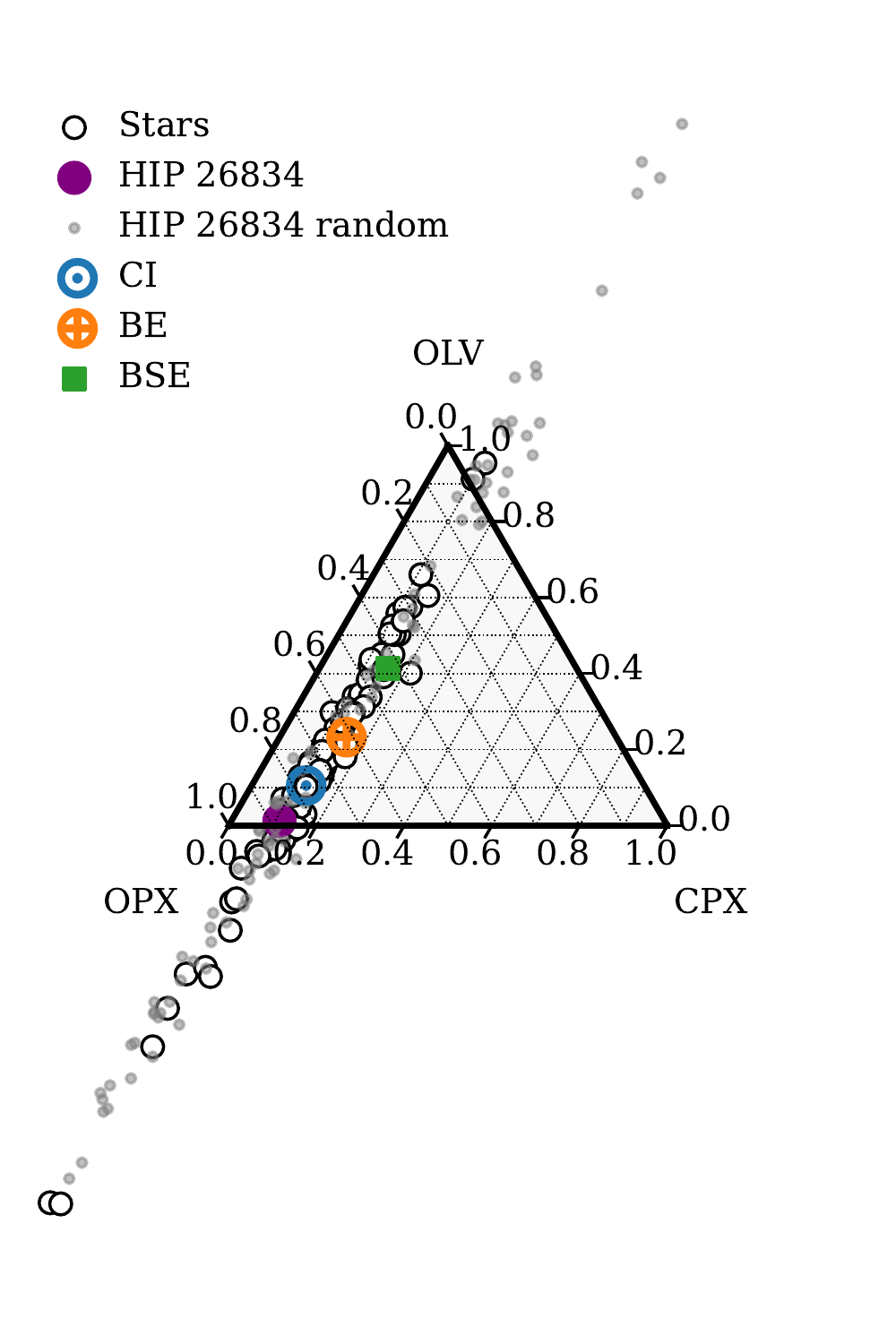} 
   \caption{Ternary diagram for 100 randomly selected Hypatia stars, and the spread in OLV, OPX, and CPX for one of the best Hypatia fits to  chondrite, HIP 26834 ($\chi^2_\nu \sim 0.02$). Despite being statistically indistinguishable from chondritic composition, the uncertainties in the measured abundances lead to huge spread in OLV and OPX quantities. Similar OLX, OPX, and CPX ranges resulting from propagation of uncertainties are found for all of the Hypatia stars. }\label{Fig:tern_best}
\end{figure}

Projecting the Hypatia catalog stellar abundances into normative mineralogy ternary space, we find, as with the WDs, the uncertainties are too large to constrain the volumetric proportions of minerals in a meaningful way. To illustrate this, Figure \ref{Fig:H_1-1} shows the abundances relative to chondrite for one of the Hypatia catalog samples, HIP 26834, yielding an excellent fit to a chondritic bulk composition. Figure \ref{Fig:H_1-1} shows that the uncertainties in abundances are relatively low for this star, but they nonetheless create a very large spread in OLV and OPX fractions (Figure \ref{Fig:tern_best}). Similar to the WDs, calculating the normative mineralogy for all of the Hypatia catalog stars results in a large spread in OLV and OPX values that reflect only uncertainties. This is consistent with \cite{Hinkel2018}, who find that much smaller measurement uncertainties than those of current observations are required to differentiate between unique planetary structures using stellar data. 

\subsection{Abundance Ratio Trends in Hypatia Catalog Stars}\label{subsection:hypatia_abundance}
The Hypatia catalog stars exhibit some systematic trends in element abundances due to galactic chemical evolution (GCE) \citep{Hinkel2014}. In particular, we note decreasing abundances of $\alpha$ elements relative to iron with increasing [Fe/H], where the latter is a non-linear proxy for time. This trend is well studied in the Milky Way and other galaxies in the local universe, and is broadly due to increased injection of Fe into the interstellar medium (ISM)  at later times due to the delayed effects of Type Ia supernovae. The late injection alters the $\alpha$ element-to-Fe ratios established by core-collapse supernovae that dominated the ISM at earlier times \citep[e.g.][]{Hayden2015, Kobayashi2020}. 

Of the elements considered in this study, Fe, Cr, and Ni abundances accelerated with time in the Galaxy as a result of late-forming Type Ia supernovae accounting for about half of their overall production. The $\alpha$ elements Mg, Si, and Ca, on the other hand, are  produced in Type II core-collapse supernovae, and increase more steadily with time in the Galaxy. Aluminum is somewhat separate from these two groups; it is also produced by Type II supernovae like the $\alpha$ elements, but the yield depends more strongly on the metallicity of progenitor stars \citep{Kobayashi2011}, and therefore exhibits a relatively small acceleration in abundance with time. The $\alpha$ elements and Al are lithophile elements while Fe, Cr, and Ni are siderophile. We note that the Hypatia catalog contains a few thousand stars in relatively close proximity to the Earth, and that trends in stellar composition therefore don't include the the wide ranges in ages or environmental affects that are observed in larger surveys \citep[e.g.][]{Horta2022}.

\begin{figure}
\centering
    \includegraphics[width=0.5\textwidth]{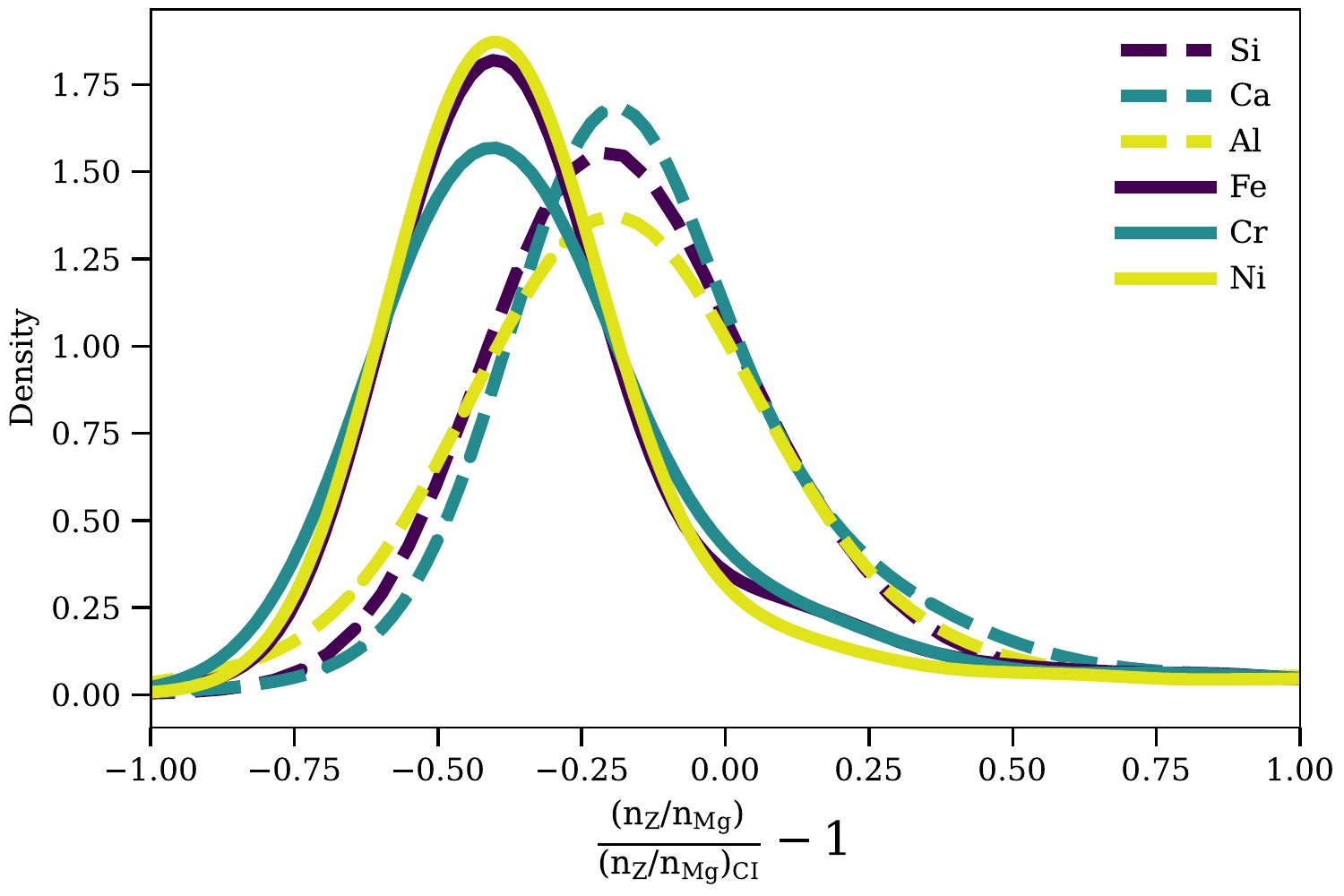} 
    \caption{The fractional difference between the measured abundances in the Hypatia catalog stars relative to chondrite, for the stars that do not pass as chondritic. For the non-chondritic stars, siderophiles (Fe, Cr, Ni) tend to be the worst fitting elements rather than lithophiles.  }\label{Fig:frac_diff_hypatia}
\end{figure}

We find that the fits to chondrite are influenced by the evolving lithophile/siderophile ratios. In Figure \ref{Fig:frac_diff_hypatia}, we show the fractional difference between the observed abundances and chondrite for the Hypatia stars that do not pass as chondritic. For the chondritic stars, all of these distributions are centered at zero. However, Figure \ref{Fig:frac_diff_hypatia} shows that Fe, Cr, and Ni abundances relative to chondrite are lower than those of the lithophile elements by about a factor of two. This suggests that Type Ia products are inflating the $\chi^2_\nu$s of non-chondritic stars relative to the $\alpha$ elements. Quantitatively, we find that of the $\sim 2000$ Hypatia catalog stars that do not pass as chondritic, $71\%$ have a siderophile element as their worst fitting abundance ratio. Of this subset of stars, $71\%$ pass as chondritic if Fe, Cr, and Ni abundances are ignored, meaning that when stars have anomalous siderophile abundances relative to the chondrite, they typically fail as chondritic because of the siderophiles. Meanwhile, $4\%$ of stars with lithophiles as the worst fitting element pass as chondritic when lithophiles are ignored. In other words, the majority of stars that fail with anomalous lithophile elements are not failing solely because of the lithophile elements. We do not see these same patterns in the WD data. Amongst the WDs, 7/15 of the failures in the raw data and 2/10 of the failures in the steady-state are due to siderophiles. We do not find higher recovery rates amongst the siderophile failures when removing siderophiles elements.

\begin{figure*}
\centering
    \includegraphics[width=0.9\textwidth]{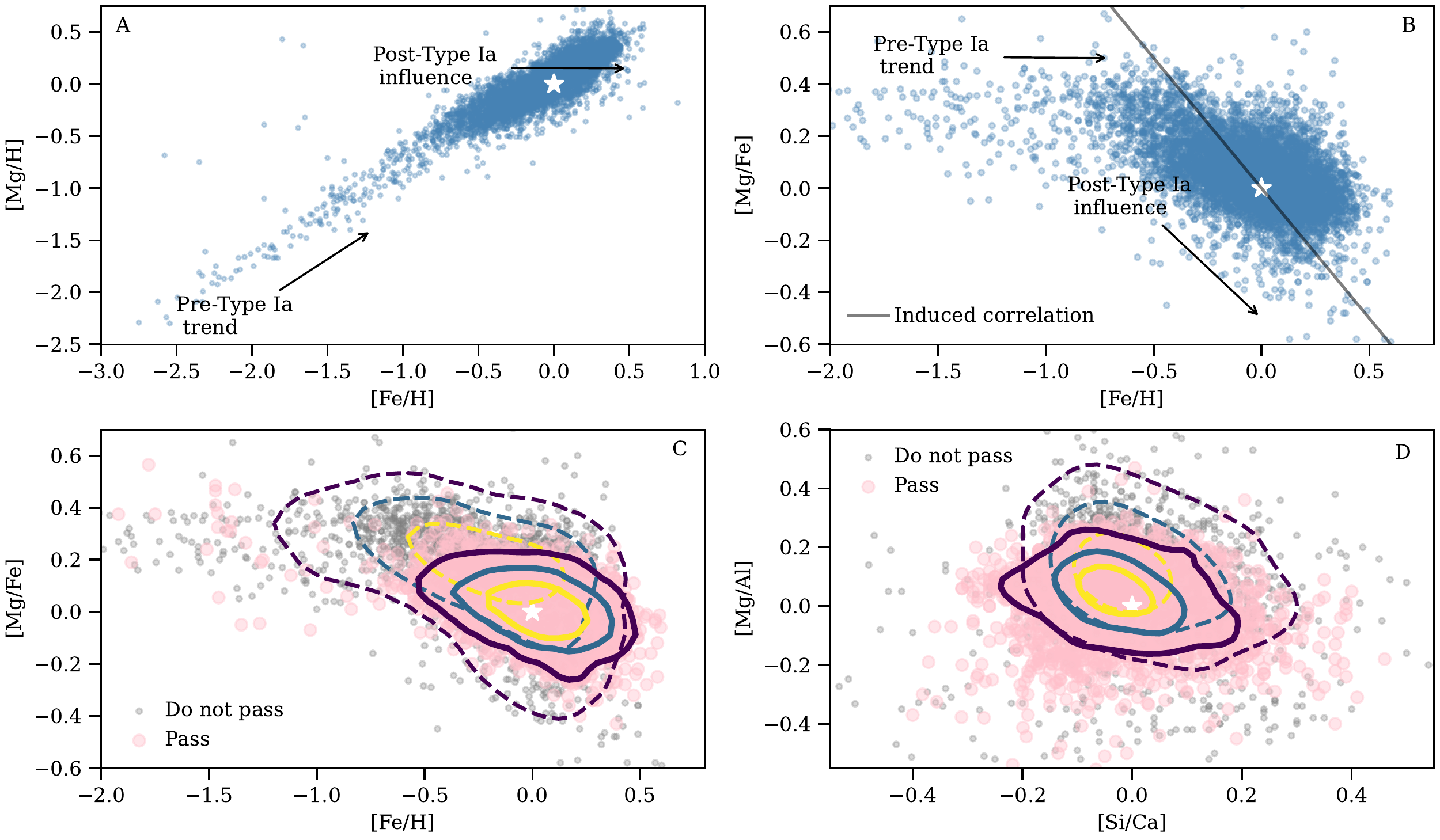} 
    \caption{Elemental abundance ratios for the Hypatia catalog stars in dex. Throughout, the white star shows solar values. A) [Mg/H] vs.\ [Fe/H], showing growth of an $\alpha$ nuclide compared with Fe. [Fe/H] is broadly taken as an indicator of time.  The influence of Type Ia supernovae on galactic chemical evolution is indicated by the two arrows. B) Same data as [Mg/Fe] vs. [Fe/H]. The solid negatively sloping line shows the effect of induced ratio correlation due to Fe appearing on both axes. C) Same as B) but with contours showing 50, 80 and 95\% levels for the chondritic (solid contours) and non-chondritic (dashed contours) populations of stars based on $\chi^2_\nu$ tests. The larger points indicate stars that pass as chondritic and the smaller points do not pass. The non-chondritic population generally extends to lower metallicities than the chondritic stars. D) Ratios of lithophile elements only, showing no clear trend.  }\label{Fig:mg_fe_ratio}
\end{figure*}

In Figure \ref{Fig:mg_fe_ratio} we show four plots of abundance ratios of the Hypatia stars, illustrating the effect of GCE on the goodness of fit to chondrite. The overall trends in relative abundances of lithophile and siderophile elements are plotted as [Mg/H] (lithophile, $\alpha$ nuclide) against [Fe/H] (siderophile, and a proxy for time) in panel A and the corresponding [Mg/Fe] ratios against [Fe/H] in panel B. As a zero-order approximation of chemical evolution in the local neighborhood, we categorize the trends in the data into two stages of pre- and post injection of Fe, Cr, and Ni by Type Ia supernovae. The break between trends is around [Fe/H] $\sim -0.5$, corresponding to $\sim 8$ billion years before present \citep{Bellardini2022}. The pre-Type Ia arrow in Figure \ref{Fig:mg_fe_ratio} shows the general trends in $\alpha$ nuclides (lithophiles), represented here by Mg, relative to siderophile abundances at low metallicities prior to the influence of Type Ia supernovae on the ISM. The post-Type Ia arrow shows the influence of Type Ia supernovae on  higher metallicity stars after the influence of Type Ia supernovae on the ISM.  The line in panel B shows the induced correlation between [Mg/Fe] and [Fe/H] that would be expected if Mg abundances were completely independent of Fe. At lower metallicity, we find that Mg and Fe abundances increase at very nearly the same rate, resulting in nearly constant [Mg/Fe] with metallicity. The increase in ISM Fe  at later times flattens the growth of Mg vs Fe, resulting in a negative slope in [Mg/Fe] with metallicity. In panel A, we fit the low and high metallicity ranges and find a slope of $0.98$ for the low end and $0.88$ for the high end, with uncertainties in the slopes of less than $0.005$. The decrease in slope is a reflection of the influence of the Type Ia supernovae at later times. For panel B, we find slopes of $-0.03$ and $-0.19$ for the low and high metallicity ranges, respectively. 

We again show the [Mg/Fe] ratio as a function of metallacity in panel C, with points colored by whether the star passes as chondritic in the $\chi^2_\nu$ tests, as well as the occurrence levels for chondritic and non-chondritic stars. The contours illustrate the somewhat different distributions of the chondritic and non-chondritic stars. We find that stars with very low metallicity, or Fe abundances, are those that are often classified as non-chondritic.

Finally, in lithophile-lithophile space (panel D of Figure \ref{Fig:mg_fe_ratio}), we find that the Hypatia catalog stars have a range of ratios centered on the Sun (the white star in Figure \ref{Fig:mg_fe_ratio}D). Consistent with GCE models, no overarching trends in ratios are seen in this case, and we find very little separation between the ratios of the Hypatia catalog stars that are considered chondritic and those of non-chondritic stars.

We conclude that older, lower metallicity stars are less likely to be consistent with a chondritic composition. In the $\chi^2_\nu$ tests, stars that are statistically distinct from chondritic more often have low Fe, Cr, and Ni compared with solar, indicating that deviations from chondritic compositions are in part attributable to the delayed effects of Type Ia supernovae. The Hypatia catalog stars are all in relative close proximity to the Sun, so while we find that the rock-forming element ratios in most of the stars are consistent with chondrites, it is possible that this conclusion would not apply to older populations of stars or stars located outside of the local disk of the Milky Way due to the effects of GCE on lithophile/siderophile ratios. 

\section{Discussion}\label{section:discussion}
A summary of the fractions of bodies that are consistent with chondritic compositions is shown in Table \ref{Table:chi2values}. The leave-out-outliers (``LOO") column includes samples that pass as chondritic using the $\chi^2_\nu$ test when an element identified as an outlier is ignored (Section \ref{section:methods}). We find that outliers do not significantly affect the fractions of stars that are consistent with chondritic composition. Ignoring outliers changes the classification from non-chondritic to chondritic for one WD, and shifts the fraction of Hypatia catalog stars consistent with chondritic composition by less than $1\%$. 

\begin{table}
\caption{Percentage of stars in the Hypatia and WDs samples that pass the $\chi^2_\nu$ test and are considered good fits to chondritic composition, using all available elements of Si, Fe, Mg, Al, Ca, Ni, and Cr. Leave-out-outliers (LOO) percentages are calculated by considering a star to be a good fit to chondrite if removing the outlier element allows the star to pass.  }
\centering
\begin{tabular}{l|c|c}
\hline
Sample       & \% Chondritic & \% Chondritic, LOO \\ \hline
Hypatia, All & 75.0         & 75.5 \\
Hypatia, $<$150 pc & 74.7        & 75.0 \\
Hypatia, F   & 66.5         & 67.0 \\
Hypatia, G   & 75.8         & 75.9 \\
Hypatia, K   & 76.6         & 76.7 \\
Hypatia, M   & 56.1         & 56.8 \\
WD, raw      & 48.4         & 51.6 \\
WD, SS       & 67.7         & 67.7 
\end{tabular}\label{Table:chi2values} 
\end{table}

\begin{figure}
\centering
    \includegraphics[width=0.5\textwidth]{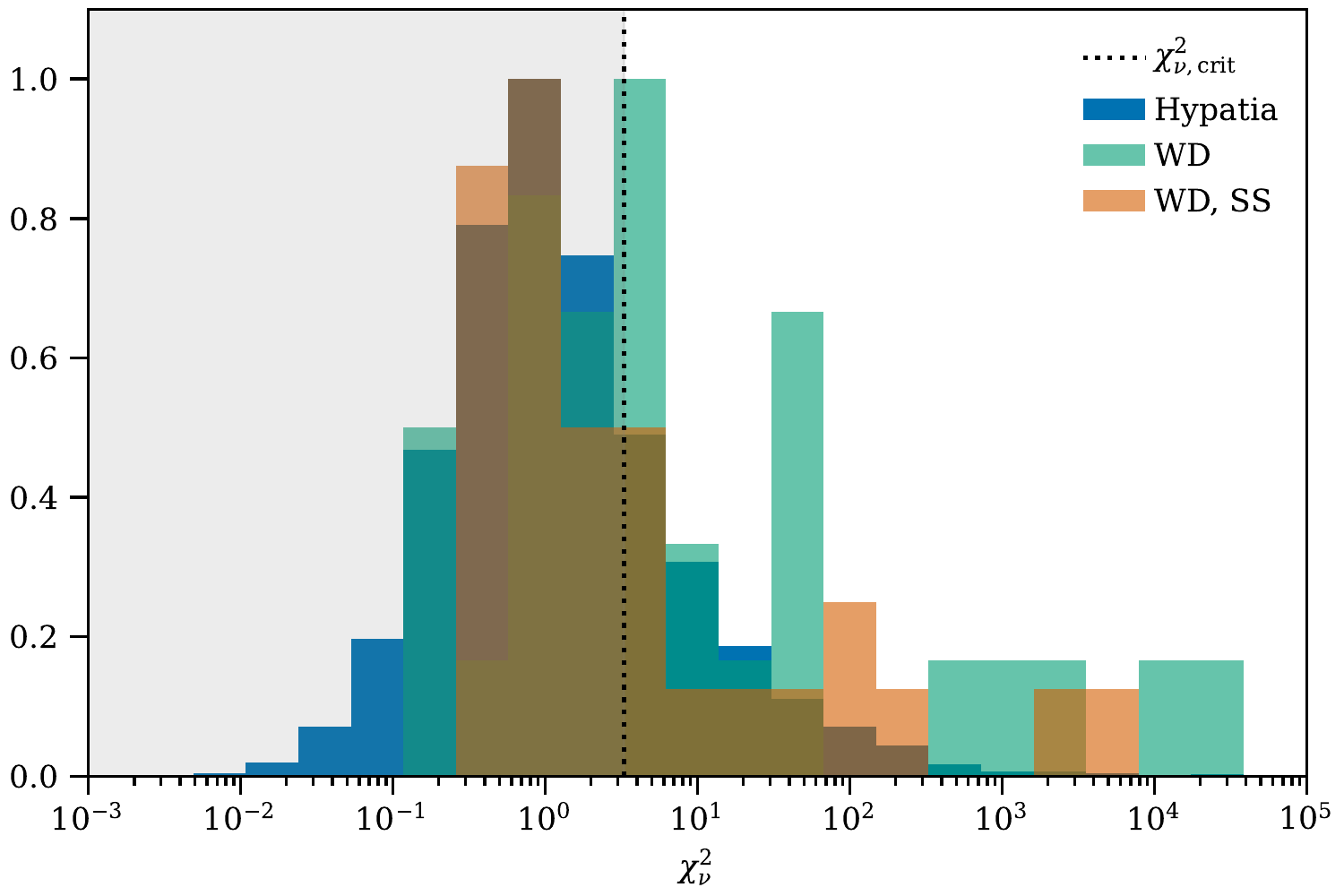} 
   \caption{Distribution of $\chi^2_\nu$ for the WDs (raw and steady-state adjusted), compared to the Hypatia catalog stars. The vertical line shows the approximate critical $\chi^2_\nu$ value ($\chi^2_\nu \sim 3$ to 4, based on the number of observed elements), so that samples within the shaded regions are considered to be consistent with a chondritic composition. }\label{Fig:H_chi2_HandWD}
\end{figure}

In Figure \ref{Fig:H_chi2_HandWD} we show the distribution of $\chi^2_\nu$ values calculated for the Hypatia catalog sample and the raw and steady-state adjusted WD data. The $\chi^2_\nu$ for all of the populations is most strongly peaked at low values, consistent with chondritic compositions. This suggests that the majority of extrasolar rocks in the solar neighborhood are built from material similar in composition to that which formed the Solar System. The overwhelming fraction of Hypatia stars with chondritic rock-forming element ratios suggests that any deviations from chondrite-like compositions observed in exoplanets are more likely to be  a result of the specific processing during planet formation  rather than the result of large differences in the initial protoplanetary source material from chondritic. M stars in the Hypatia data set exhibit a tail to higher $\chi^2_\nu$ values, though the majority of M stars still pass as chondritic. The difference in the M dwarf distribution relative to the others is evidently a result of different treatments of errors at near and far distances (M dwarfs are nearer) and potential systematic offsets in Ca. 

Many of the WDs and Hypatia catalog stars that did not pass as chondritic have high relative Mg concentrations; their abundance ratios in Figures \ref{Fig:O_WD_1-1} and \ref{Fig:H_1-1} (lower panel) all fall below the 1-1 line for chondritic composition due to an excess in the  Mg concentration used as the denominator in all ratios. Because we ratio to Mg, high Mg can systematically draw  chondritic abundance ratios away from chondrite, inflating the $\chi^2_\nu$ values. Examples amongst the WDs (raw data) include WD1415+234, SDSSJ2339-0424, SDSSJ1242+5526, WD1232+563, SDSSJ0738+1835, WD1350-162, and WD1929+012. If outliers are ignored, then G241-6 and SDSSJ1043+0855 are also fall into this list. For the WDs, applying the steady-state adjustment brings some, though not all, of the elements from the apparently Mg-rich WDs back to or above chondritic abundances. We find, therefore, that for WDs with excess Mg,  deviations from chondritic in these cases is due in no small measure to the effects of settling. 

We now explore how this study of local stars and WDs fits in to both the overarching metallicity gradients in the Galaxy and the current landscape of inferred exoplanet compositions. 

\subsection{Galactic Chemical Evolution}\label{section:GCE}
The Milky Way experiences spatial and temporal variations in stellar compositions, begging the question of how representative the pervasive chondritic compositions we see in the solar neighborhood are with respect to time and place in the Galaxy. In Section \ref{subsection:hypatia_abundance} we showed that galactic chemical evolution (GCE) has implications for the relative lithophile to siderophile ratios in stars over time, though the bulk of the Hypatia catalog stars are still consistent with chondrites. Here we evaluate the significance of chondritic rock-forming element ratios in the context of large scale variability in the Galaxy, outside of the solar neighborhood. 

Spatial metallicity gradients in the Milky Way exist both radially and vertically (Galactic latitude) as a result of GCE. The disk midplane tends to have more metal-rich stars than above or below the plane and the disk itself exhibits a negative gradient, with generally higher metallicities towards the galactic center \citep[e.g.][]{Hayden2014, Donor2020}. Radial compositional changes may arise as annuli of the Milky Way are differentially enriched by supernovae and stellar feedback. For example, \cite{Bellardini2022} find from cosmological simulations that the older, inner disk receives more material from Type Ia supernovae, leading to lower [Mg/Fe]  compared to the outer disk. They also find that some azimuthal scatter in abundances is to be expected, though the scatter is relatively low ($<\sim 0.05 \rm dex$). While variations in [Fe/H] of $> 1$ dex are found across the entire galactic disk, much smaller variations in [Fe/H] of about $\pm 0.2$ dex are found for stars within 2-3 kpc of the Sun \citep{Monson2017}. Radial variations in metallicity may be further damped by radial migration and mixing of stars throughout the disk.

Overall metallicity is expected to rise with time in the Galaxy.  For example \cite{Timmes1995} showed that in the solar neighborhood at galactocentric radii of about 8 kpc, changes in [Fe/H] of about 1.5 dex are to be expected over 14 Gyr. However, the majority of this increase in metallicity occurs within the first few Gyr of galactic evolution, with changes of less than 0.5 dex [Fe/H] from about 2 Gyr onwards. 

Therefore, while significant compositional changes occurred very early in the Milky Way's evolution, or very close to the galactic center, we do not expect to find demonstrable effects of GCE in rock-forming element ratios among stars in the stellar disk at galactocentric radii between $\sim 4$ kpc and $\sim 10$ kpc as seen today. 

Given these trends in GCE, we now assess the impact on polluted WDs by estimating their formation times. The polluted WDs in our sample have cooling ages of about $50-600$ Myr and masses between $\sim 0.5-0.75 M_\odot$ (Table \ref{Table:star_table}). These WD masses translate to initial stellar masses of $\sim 1-3 M_\odot$ \citep{Cummings2018, El-Badry2018} and stellar lifetimes of $\rm \sim 300 Myr - 10 Gyr$ \citep{Schaller1992}. Meanwhile, radioactive dating has constrained the age of the Milky Way to $\sim 13.8 \rm Gyr$ \citep[e.g.][]{Cayrel2001, Hill2002, Cowan2002}. Given overall metallicity changes are most significant in the first few Gyr of the Galaxy, this suggests that GCE trends could only manifest in the lowest-mass WD in our data set (GaiaJ0218+3625), with progenitor masses approximately that of the Sun, and corresponding lifetimes exceeding 10 Gyr. Within our sample, we do not find evidence that lower mass WDs are worse fits to chondrite, however it is possible that such trends may be evident in future, larger samples of WDs.

We conclude that GCE could significantly alter the relative abundances of rock-forming material available for planet formation at early times and outside of the disk of the Milky Way. We explore a possible implication of this in the next section. We find that the majority of the current population of polluted WDs are derived from sufficiently young progenitors and nearby to the Sun that we do not expect GCE to have a strong affect on their compositions.  

\subsection{Iron core mass fractions}
One of the possible consequences of GCE for exoplanets is the resulting core-mass fractions due to variations in lithophile/siderophile ratios. The mass fractions of iron-rich cores of rocky planets have been shown to be related to the iron mass fractions deduced from their host stars \citep[e.g.,][]{Adibekyan2021, Rogers2021}. Consistent with these studies, we calculate the iron mass fractions of planets that might have formed from the material polluting WDs or around the Hypatia catalog stars as  $f_{\rm Fe} = m_{\rm Fe} / \left(m_{\rm Fe} + m_{\rm Mg_2SiO_4} + m_{\rm MgSiO_3} + m_{\rm SiO_2}\right)$, where $m_i$ is the abundance of the species $i$ relative to H or He  multiplied by the formula weight \citep{Santos2015}. The relative abundances by number of the silicate species are obtained from a linear transformation such that MgSiO$_3 =$2Si$-$Mg, Mg$_2$SiO$_4=$ Mg$-$Si, and SiO$_2$ is any remaining Si. The mass of O in the rock is therefore derived from the Si and Mg abundances, which corrects for any O excesses due to water (for the WD sample) or O production in the Hypatia catalog stars. 

\begin{figure}
\centering
    \includegraphics[width=0.5\textwidth]{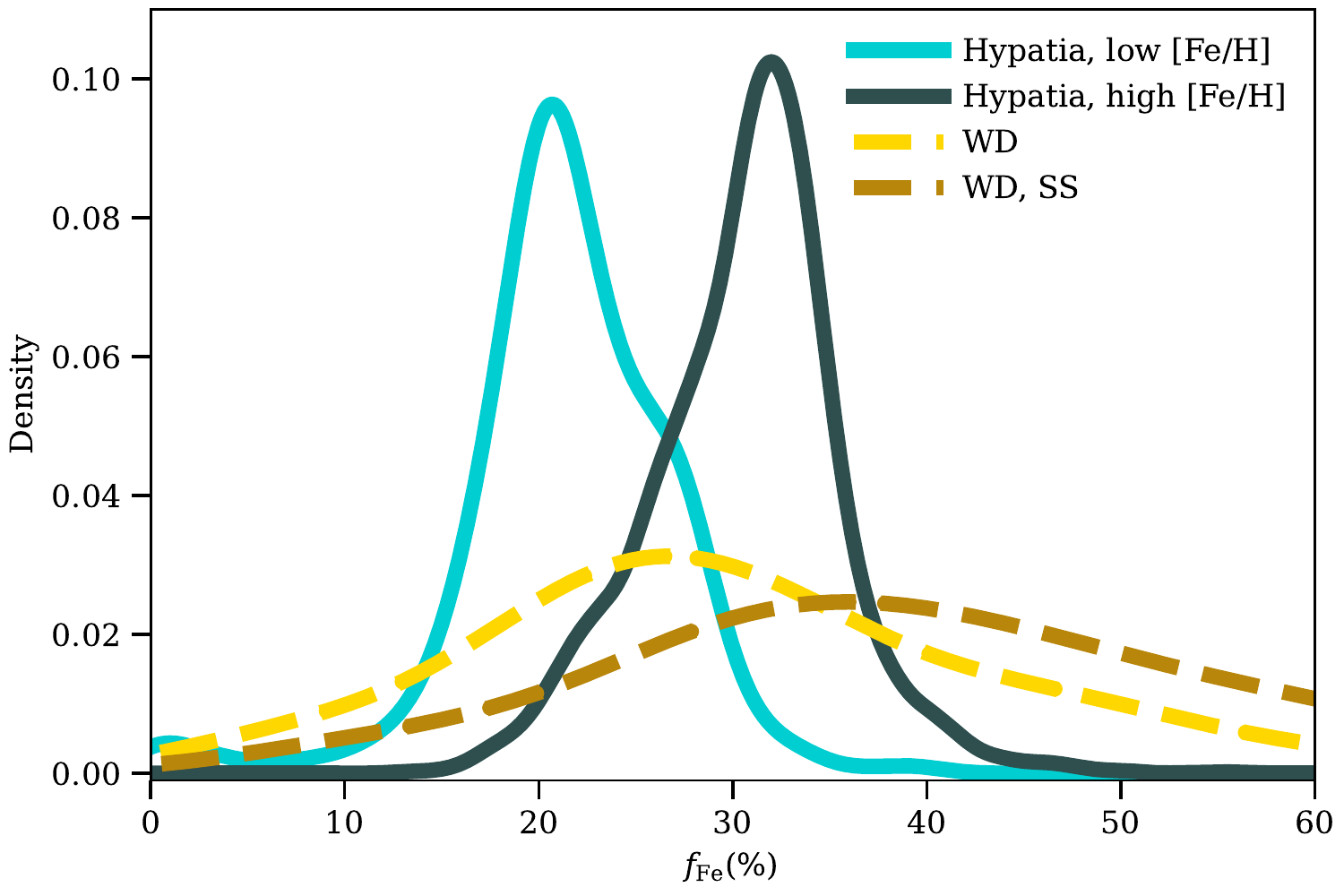} 
    \includegraphics[width=0.5\textwidth]{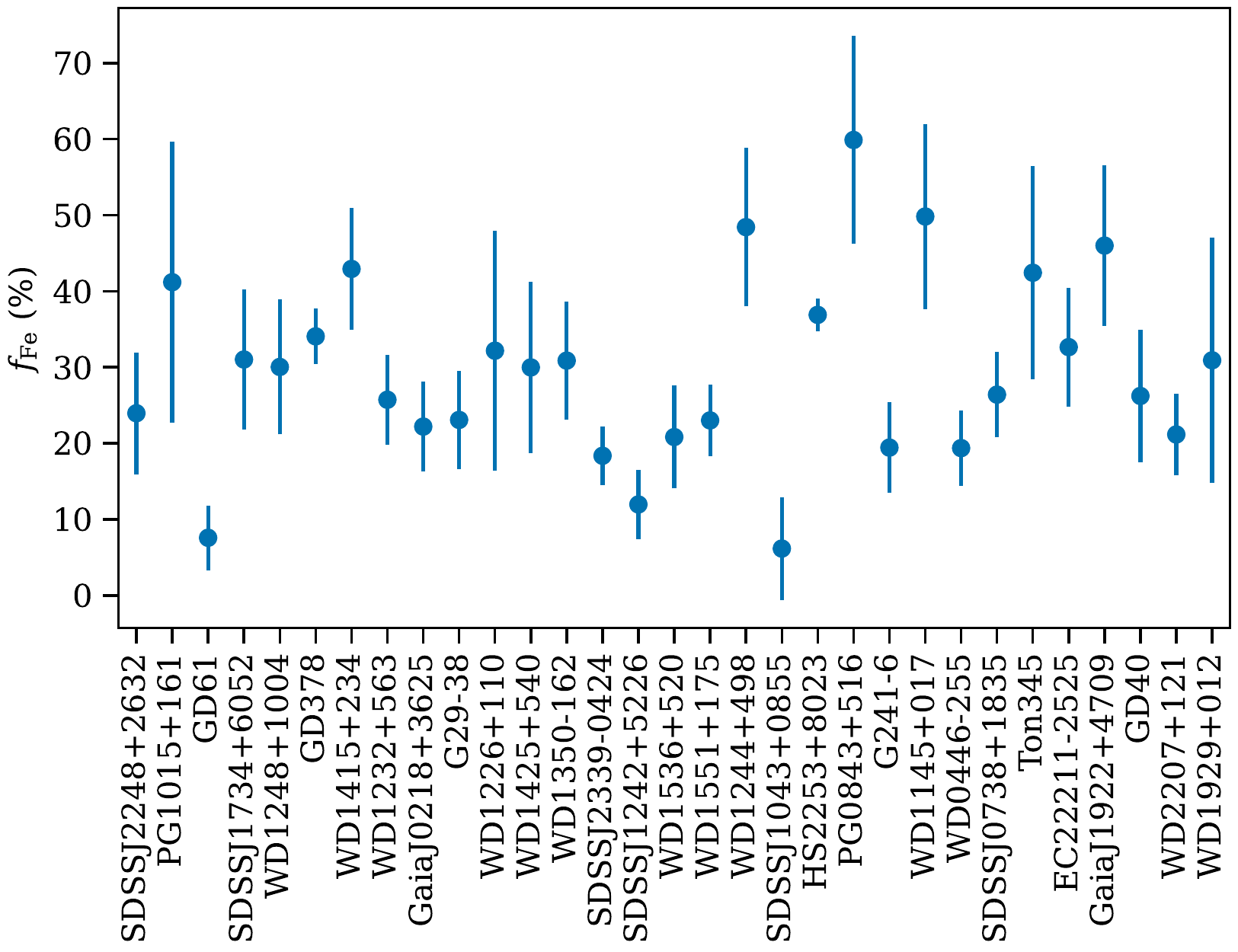} 
   \caption{Top: Distribution of iron mass fractions $(f_{\rm Fe})$ for the WDs and Hypatia 
 catalog stars. We show both raw and steady-state values for the WDs, and split the Hypatia catalog stars into low ([Fe/H]$<-0.5$) and high ([Fe/H]$>-0.5$) metallicity categories.  Bottom: Median and $1\sigma$ spread in $f_{\rm Fe}$ for the sample of WDs based on 100 random draws from their median and uncertainty in the raw Si, Mg, and Fe abundances. Most WD iron fractions have a range of at least $10\%$ when accounting for uncertainty in their abundances. }\label{Fig:fFe}
\end{figure}

The mass fractions of Fe can be equated with the metal core fractions of planets given the expectation of small concentrations of Fe in the silicate \citep{Doyle2019}. The top panel of Figure \ref{Fig:fFe} shows $f_{\rm Fe}$ calculated from the abundances of the raw and steady-state adjusted WD data, and the Hypatia stellar abundances. The bottom panel illustrates the distribution of $f_{\rm Fe}$ resulting for four example WDs when uncertainties in observation are propagated through the transformation. 

 The  Hypatia catalog stars with higher metallicities define a slightly skewed distribution of iron mass fractions, with a well-defined mode of about $32\%$, indistinguishable from the core mass fraction of the Earth, a tail towards lower values, and an approximate $1 \sigma$ spread of about $\pm 5 \%$. The lower metallicity Hypatia catalog stars define a peak in the distribution of $f_{\rm Fe} \sim 20 \pm 5 \%$ (Figure \ref{Fig:fFe}). A similar variation in iron mass fractions was calculated by \cite{Michel2020} for stars in the thin and thick disks and halo of the galaxy.
 
 The difference in most probable iron mass fractions obtained from the higher and lower metallicity Hypatia catalog stars suggests the possibility that planets formed in first several billion years in the Milky Way may have tended to have smaller metal cores compared with Earth, while planets formed later are generally similar to Earth in their metal core fractions. This is broadly due to the increase in siderophile elements at later times relative to lithophiles. 
 
 The WD sample size is much smaller, and plagued by larger uncertainties. The bottom panel of Figure \ref{Fig:fFe} shows the spread in $f_{\rm Fe}$ obtained for each WD after taking 100 random draws of of Si, Mg, and Fe abundances from a parent population defined by the WD medians and uncertainties in each element. The $1\sigma$ uncertainty in iron mass fraction ranges from about $4-15\%$, with a median of $\sim7\%$. In any case, the raw data define iron mass fractions peaking at $20-30\%$ while the steady-state adjusted data yield a peak at $30-40\%$. While we do calculate low ($f_{\rm Fe}<10\%$) iron mass fractions for some WDs, lacking further information about the ages or initial metallicities of these stars prevents us from identifying whether the low iron is due to the systematic effects of GCE. As a whole, these data suggest the majority of planets that might have formed from these polluting materials have metal core mass fractions that are not significantly different from an Earth-like planet. Following the discussion in Section \ref{section:GCE}, abundance measurements for older WDs (likely lower-mass WDs), or WDs outside of the thin disk of the galaxy could help identify whether the lower core mass fractions are influenced by GCE. 

\section{Conclusions}\label{section:conclusion}
In this work, we show that about half of polluted WDs, and well over half of the Hypatia catalog stars, have compositions that are consistent with chondrites. We use the $\chi^2_\nu$ goodness of fit to test the composition of each star, with a threshold of $\alpha =0.05$ to select matches to chondritic composition, and allow for a $2\sigma$ error in the $\chi^2_\nu$ to account for our small sample size of observed elements. 

We use Monte Carlo methods to propagate the uncertainties in the observed abundances of the WDs and Hypatia catalog stars. We find that many Solar System rocks, including bulk Earth and bulk silicate Earth, bulk silicate Mars, and E chondrites, are indistinguishable from CI chondrite given current uncertainties in WD pollution measurements. Additionally, we find that we are not able to characterize either Hypatia catalog stellar abundances or WD pollution by normative mineralogies  due to the impossibly large uncertainties obtained by propagating measurement uncertainties.  

The polluted WD data indicate that the bulk of exo-rocks are consistent with chondritic compositions. This is supported by the compositions of rocks implied by the Hypatia catalog stars, which suggest most material in the solar neighborhood formed in protoplanetary disks with rock-forming element ratios similar to our Sun. The Hypatia catalog stars do suggest, however, that galactic chemical evolution can lead to exoplanet compositions statistically different from the Solar System in the first few billion years of the Galaxy or in galactic substructures with considerably different metallicities. One implication of this is that earlier in the evolution of the Milky Way, rocky planets may have formed with substantially less massive metal cores than Earth. 

Our methods do not suggest any of the WD polluters are composed of crust, either MORB or continental crust. No stars in our sample are better fits to MORB or continental crust than chondrite, even WDs with the largest deviations from chondritic composition ($\chi^2_\nu \gg 10$). We conclude that the relative abundances of rock-forming elements in polluted WDs and local stars are relatively homogeneous, which suggests that the majority of extrasolar rocks in the solar neighborhood originate from chondrite-like compositions. 

\acknowledgements
The authors thank Pratik Gandhi (University of California, Davis) for helpful discussions on galactic chemical evolution. We also thank the referees for their comments, which improved the manuscript. This work was supported by NASA 2XRP grant No. 80NSSC20K0270 to EDY. 

The research shown here acknowledges use of the Hypatia catalog Database, an online compilation of stellar abundance data as described in \cite{Hinkel2014}, which was supported by NASA's Nexus for Exoplanet System Science (NExSS) research coordination network and the Vanderbilt Initiative in Data-Intensive Astrophysics (VIDA).


\bibliography{bibli}{}
\bibliographystyle{aasjournal}



\end{document}